\documentclass[namedreferences]{SolarPhysics}
\usepackage[optionalrh]{spr-sola-addons} 
\usepackage{graphicx}        
\usepackage{amssymb}        
\usepackage{color}           
\usepackage{url}             

\newcommand{\aap}{    {\it Astron. Astrophys.}}

\newcommand{\apj}{    {\it Astrophys. J.}}

\newcommand{\solphys}{{\it Solar Phys.}}

\begin{document}

\begin{article}

\begin{opening}

\title{3D evolution of a filament disappearance event observed by STEREO}

\author{S.~\surname{Gosain}$^{1}$\sep
        B.~\surname{Schmieder}$^{2}$\sep
        P.~\surname{Venkatakrishnan}$^{1}$\sep
        R.~\surname{Chandra}$^{2}$\sep
        G.~\surname{Artzner}$^{3}$
       }
\runningauthor{Gosain et al.}
\runningtitle{Disappearing filament seen with STEREO}

   \institute{$^{1}$ Udaipur Solar Observatory, P. Box 198, Dewali, Udaipur 313001
                     email: \url{sgosain@prl.res.in} \\
                     email: \url{pvk@prl.res.in} \\
              $^{2}$ Observatoire de Paris, LESIA, 92190 Meudon, France
                      email: \url{Brigitte.Schmieder@obspm.fr} \\
                      email: \url{Chandra.Ramesh@obspm.fr} \\
              $^{3}$ Institut d' Astrophysique Spatiale Bat. 121 91405, Orsay, France
                     email: \url{Guy.Artzner@ias.u-psud.fr} \\
             }

\begin{abstract}
A filament disappearance event was observed  on 22 May 2008  during our recent
campaign JOP 178. The filament,
situated in the southern hemisphere, showed sinistral chirality
consistent with the hemispheric rule.  The
event was well observed by several observatories in particular by
THEMIS.   One day before the disappearance, H$\alpha$ observations showed up and down flows in adjacent locations along the filament, which suggest plasma motions along twisted flux rope. THEMIS and GONG  observations show shearing photospheric motions leading to magnetic flux canceling around barbs. STEREO A, B spacecraft with separation angle 52.4 degrees, showed quite different views of this
 untwisting flux rope in He II 304 \AA\ images.  Here, we reconstruct the 3D geometry of the filament during its eruption phase
using STEREO EUV He II 304 \AA\ images and find that the filament was highly inclined to the solar normal.
The He II 304 \AA\ movies show individual threads, which oscillate and rise  to an altitude of about 120 Mm  with apparent velocities of about 100 km s$^{-1}$, during the rapid evolution phase. Finally, as the flux rope expands into the corona, the filament disappears by becoming optically thin to undetectable levels. No CME was detected by STEREO, only a faint CME was recorded by LASCO at the beginning of the disappearance phase at 02:00 UT, which could be due to partial filament eruption.  Further,  STEREO Fe XII 195 \AA\ images showed bright loops beneath the filament prior to the disappearance phase, suggesting  magnetic reconnection below the flux rope.
\end{abstract}
\end{opening}
\section{Introduction}
     \label{S-Introduction}
  Two types of structures are commonly used to model filaments i.e., arcades structure or twisted flux tubes. Evidence of twisted flux tubes clearly appear during the
eruptive phase of prominences \cite{Gary2004},
\cite{Torok2005}. Most quantitative models assume that the flux rope
is in static equilibrium. 3-D magnetic models of filaments by
extrapolating photospheric magnetograms into the corona have been
developed \cite{1998A&A...329.1125A}, \cite{Aulanier2000}, \cite{Dudik2008}, \cite{aad2004}. Such
models reproduce helical ropes overlying the polarity inversion line (PIL) and the filament plasma
is assumed to be located in  dips of the helical field lines. In areas
where magnetic parasitic polarity elements are located close to the PIL,
the dips extend away from the main body of the filament creating
barbs. It is consistent with the findings of the observers
\cite{Martin94},\cite{Martin98} and \cite{Wang2001}.
 One of the possible causes of filament eruption is instability
\cite{Forbes1991},\cite{Isenberg1993}.
Determination of true filament height is an important parameter in
studies of filament eruption \cite{Schrijver2008}.

Traditionally, this has been a difficult task and only via
H$\alpha$ observations of the limb prominence one could measure the
filament height. The disadvantage of this method being: (i) it measures only
the projected height, (ii) non-availability of magnetograms due to its
location at the limb,  (iii) can not be used to follow height
evolution for several days and (iv) no continuity of multi-temperature observations.
Only recently, with the advent of stereoscopic observations by the
twin spacecraft of the {\it Solar Terrestrial Relations Observatory} (STEREO)
mission called STEREO-A (Ahead) and STEREO-B (Behind), the true
height of filament can be judged properly and the heating of the plasma can be tested
\cite{2008SSRv..136....5K};\cite{2008SoPh..252..397G};\cite{Liewer09}.

   Here in this paper, we report on the multi-wavelength observations
of a filament
   eruption event using ground as well as space based observatories
during a joint observing campaign (JOP-178 from 20 to 25 May 2008).
   The filament was located in a large filament channel with very weak
and diffuse polarities. The weak magnetic polarities in
   the filament channel are recognized using THEMIS/MTR instrument
which has high polarimetric sensitivity and a simultaneous H$\alpha$
scan.
   The evolution of these polarities is studied using full-disk
GONG line-of-sight magnetograms which are available at a cadence of one minute.
    Further, we reconstruct the true filament height and eruption
   velocity using stereoscopic images by SECCHI/EUV instrument  in He II
304 \AA\ ($\sim$60-80 $\times 10^{3}$ K).
   The He II 304 \AA\ observations are very useful in tracing filaments
because (i) the filament spine is much sharper and clearer
    \cite{2007AAS...21012006M};\cite{2007BASI...35..447J}, and (ii)
the filament can be traced up to higher altitudes compared to H$\alpha$ images
    \cite{2007BASI...35..447J}.  With the help of stereoscopic
observations by STEREO mission we reconstruct
the filament geometry and  height during the phase.  It is difficult to derive the rise velocity by height-time profile as the filament becomes very diffuse during the disappearance phase, however, we try to
estimate the projected rise velocity of individual threads from movies. Also, we identify possible reconnected loops below the flux rope.

\section{Observations and Results}
\begin{figure}    
\centerline{\includegraphics[width=.7\textwidth]{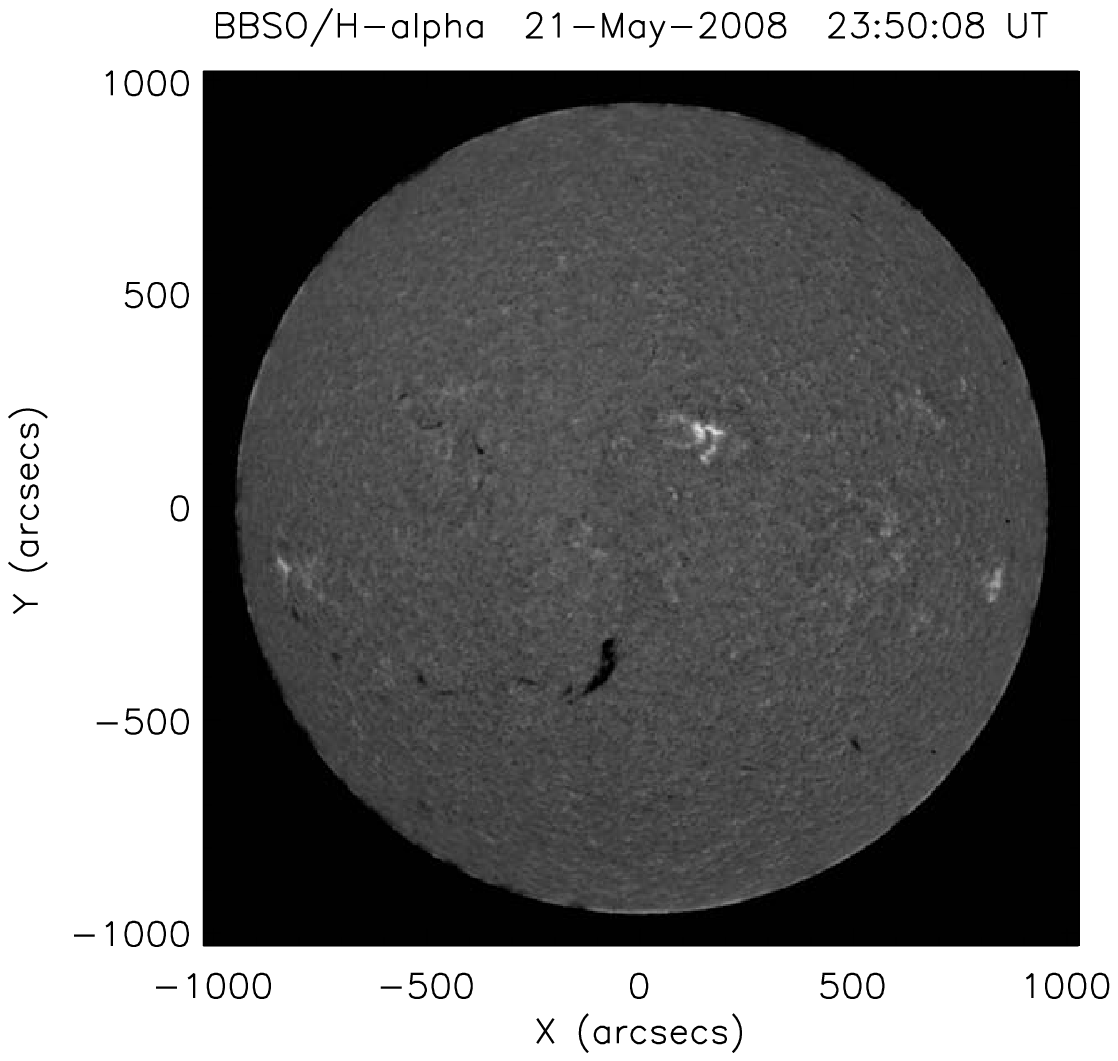}\includegraphics[width=.4\textwidth]{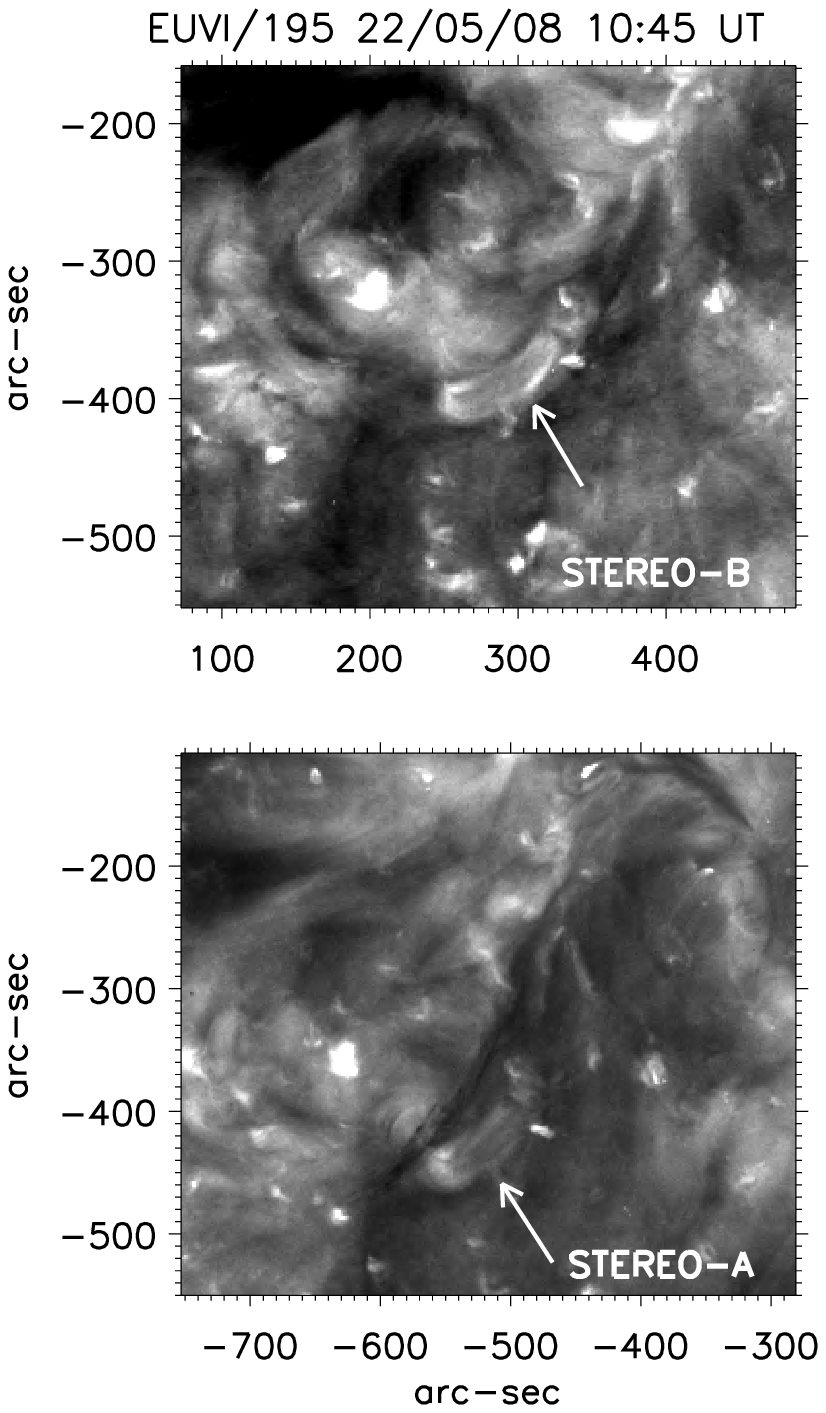}}
\vspace{0.1in}
\centerline{\includegraphics[width=1.\textwidth]{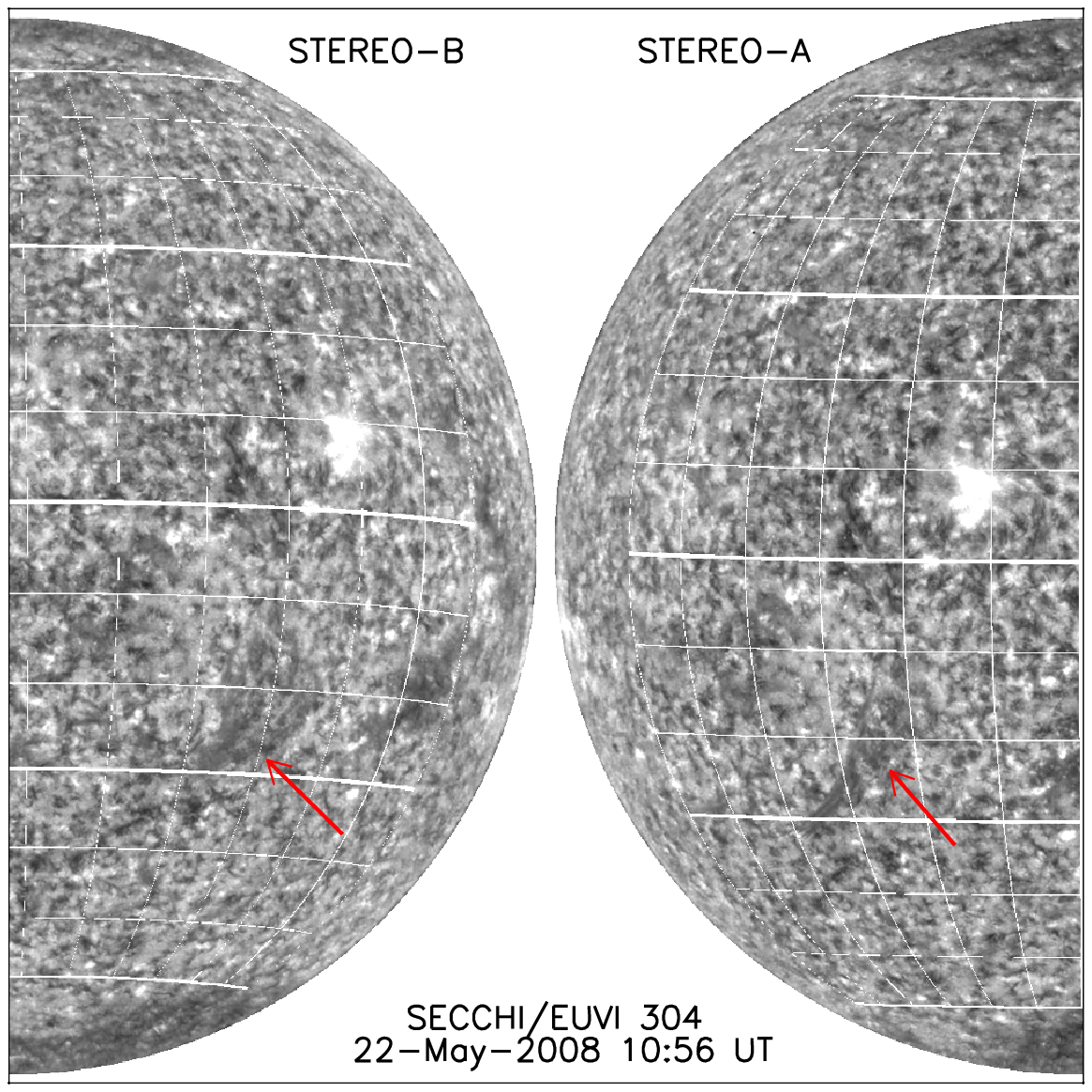}}
\caption{ The filament as seen in BBSO H$\alpha$ full-disk image on May 21 at 23:50 UT ({\it top left}) and  SECCHI/EUVI 195 \AA\ in STEREO-A and B views at 10:45 UT ({\it top right panel}). Arrows mark the bright loops discussed in the text, which are seen on either side of the filament in A and B views. Global view of the sun and filament during its erupting phase on May 22 at 10:56 UT in He II 304~\AA\, observed by STEREO A (bottom right) and B (bottom left). The images are  co-centered, re-scaled to same size and rotated to keep solar north up. The arrows indicate the filament in the two images. }
\label{fig:stereofulldisk}
\end{figure}

\begin{figure}    
\centerline{\includegraphics[width=1.0\textwidth]{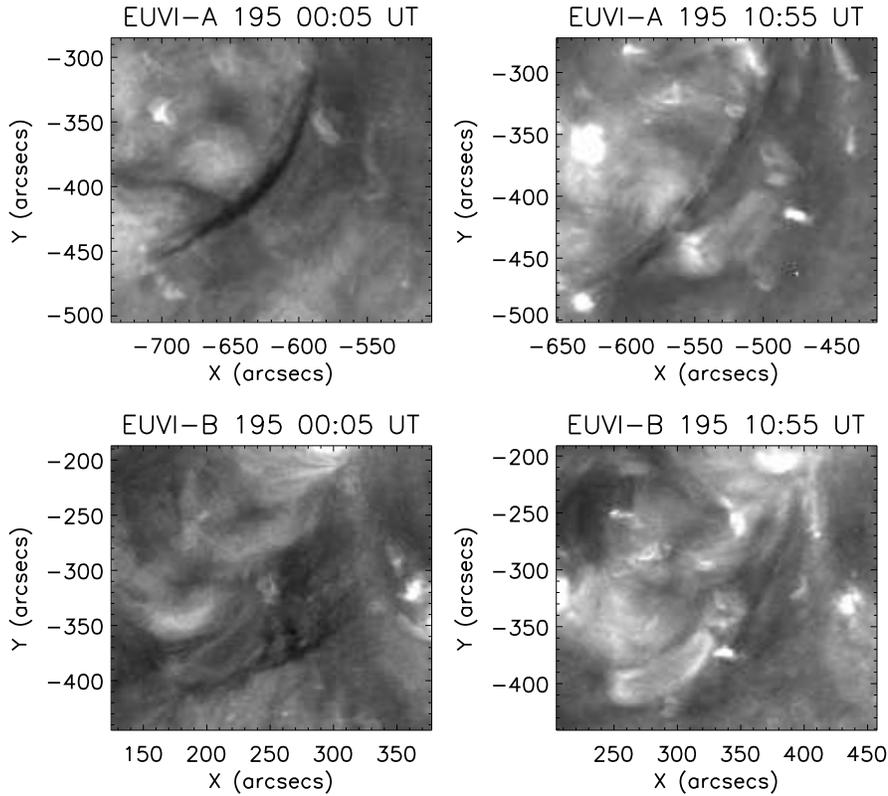}}
\caption{ The filament as seen in  Fe XII 195 \AA\ images by STEREO/EUVI.  Top row shows STEREO-A view on 22 May 2008 at 00:05 UT (left panel) and 10:55 UT (right panel). Bottom row shows STEREO-B view of the filament at the  same times. The bright loops (banana like structure) seen in EUV Fe XII 195 images are possibly the  newly reconnected loops below the filament which are not seen at 00:05 UT and clearly seen at 10:55 UT. They appear around 06:00 UT. }
\label{fig:euv195304}
\end{figure}

The JOP 178 observing campaign was carried out during 20-25 May 2008.  The observations were targeted on a filament near the polarity inversion line (PIL) in a weak bipolar magnetic region. The region was situated  in the southern hemisphere, located at 30$^\circ$~S, 30$^\circ$~E (on 20 May 2008). In Figure~\ref{fig:stereofulldisk},  a BBSO H$\alpha$ full-disk context image on 21 May, together with two views of the filament in He II 304 \AA\ by STEREO,  is shown. It may be noticed that the filament extends over a large area on the sun. The evolution of the filament was observed extensively using ground based H$\alpha$ observations also. The Table 1 gives a listing of the various ground based observations relevant to filament disappearance during JOP-178, May 20-25, 2008. These are described in the subsequent sections.  During the eruption, which took place on May 22 around 10:45 UT, a bright structure appears in the filament channel, as shown by arrow in top-right panels in Figure~\ref{fig:stereofulldisk}. This feature is not visible earlier at 00:05 UT and appears around 06:00 UT. Just few hours before the disappearance phase it is seen prominently at 10:55 UT as shown in  Figure~\ref{fig:euv195304}. A movie showing  evolution of the filament in EUV 195 \AA\ wavelength by both STEREO A and B is available in the electronic version [see, movie-195 on {\it http://tinyurl.com/movie-html}]. The question is: is it due to  heating of the filament plasma? Does this structure  correspond to overlying loops or to reconnected loops below the filament-flux rope after reconnection, like post flare loops in flares. The geometry  and the inclination of the filament  computed in section 2.3.4  leads to the third solution. This structure  is visible on the left side  of PIL with STEREO B and on the right side of STEREO A,  this implies that the structure should be located below the filament and might correspond to sheared reconnected loops.  The plasma of the filament itself is not detectable at this temperature.

 Further, two views  of the filament from widely separated vantage points, observed by STEREO using He II 304 ~\AA\ images, are shown in Figure~\ref{fig:stereofulldisk}. The separation angle of the twin spacecraft was about 52.4 degrees during our observations. The images were recorded by SECCHI/EUVI instrument at a cadence of 10 minutes. The STEREO data was reduced using SolarSoft and FESTIVAL libraries under IDL data analysis package. We define various data reduction steps that were followed before our analysis:

\renewcommand{\labelenumi}{(\alph{enumi})}

\begin{enumerate}
\item co-center the two images,
\item scale STEREO-B image to STEREO-A image size,
\item rotate image B to same roll parameters as image A,
\item orient both images to keep solar north up and overlay Carrington grid. This is useful for comparisons with Earth view.
or,
\item do not orient images for solar north up and overlay spherical grid with diameter equal to diameter of STEREO-A image. This gives epipolar view of the two images with homologous features lying along same line.
\end{enumerate}

\renewcommand{\labelenumi}{(\roman{enumi})}

In section 2.3 we present analysis of these STEREO images with  three different approaches:
\begin{enumerate}
\item  Using Movies: visually by looking at the movies of the disappearing filament. The movies were made using FESTIVAL software package \cite{Auchere08}. This involved reduction steps (a), (b), (c) and (d) above.

\item  Using Grid: by overlaying a spherical grid over the images to identify elevated structures. The data is reduced following steps (a), (b), (c) and (e).

\item  Using SCC\_MEASURE:  In order to reconstruct 3D coordinates of filament using  SCC\_MEASURE routine of the SolarSoft SECCHI library. This routine uses so called ``tiepointing" technique, where same feature is manually located in both images to reconstruct the 3D coordinates of the feature \cite{Thompson06}. The technique uses ``epipolar constraint" to reduce 2D problem to 1D \cite{Inhester06}. This approach requires data reduction steps (a), (b), (c) and step (e) without overlaying grids.
\end{enumerate}

\begin{table}
\caption{JOP-178 observations of filament during 20-25 May 2008.}
\label{tab:1}
\begin{tabular}{lcccr}
\hline
Observations &  Date  &  Time & wavelength &FOV \\
\hline
MSDP &  20 May 2008 &  11:28 to 11:54 UT&  $H\alpha$ & 5$'\times 7.5'$ \\
& 21 May 2008 &07:47 to 08:46 UT& $H\alpha$& 5$'\times 7.5'$ \\
&  & 08:51 to 09:16 UT& $H\alpha$& 5$'\times 7.5'$ \\
&  &  09:18 to 09:43 UT &$H\alpha$ & 5$'\times 7.5'$ \\
USO & 20 May 2008  & 05:36 to 10:51 UT & $H\alpha$ & 6$'\times 3'$ \\
&  21 May 2008 &  05:13 to 11:00 UT &$H\alpha$ & 6$'\times 3'$ \\
&  22 May 2008 &  05:14 to 11:00 UT &$H\alpha$ & 6$'\times 3'$ \\
MTR&  20 May 2008 &  Seq. No. 4, 14, 16 & $H\alpha$, 589.6,& $155''\times84''$ \\
&&(09:53, 15:43, 18:12 UT)& 525.0, 610.3 nm& \\
&  21 May 2008 &  Seq. No. 3, 16 &$H\alpha$, 589.6, & $45''\times84''$ \\
&& (08:38, 18:12 UT)&525.0, 610.3 nm& \\
&  22 May 2008 &  Seq. No. 2 & $H\alpha$, 589.6,& $45''\times84''$ \\
&&(09:47 UT)&525.0, 610.3 nm&\\
\hline
\end{tabular}
\end{table}

\begin{figure}    
\centerline{\includegraphics[width=1.05\textwidth,clip=]{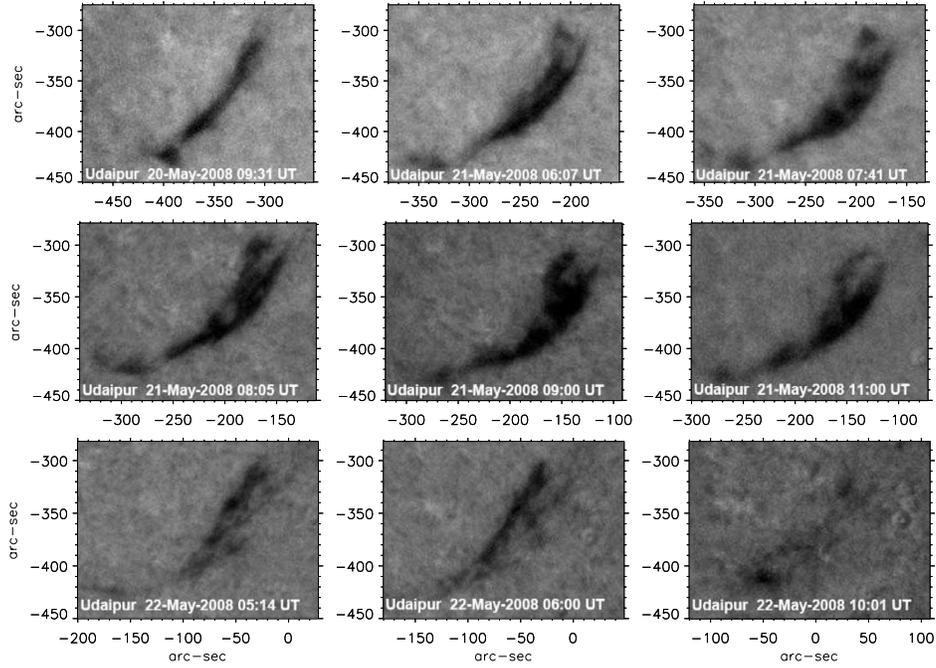}}
\caption{  The evolution of the filament from 20 to 22 May 2008, as seen in H$\alpha$ filtergrams from Udaipur Solar Observatory. The date and time of the images are mentioned in the bottom of the corresponding panel. The filament is vanishing on May 22 (see bottom right image).}
\label{fig:uso}
\end{figure}

\begin{figure}    
\centerline{\includegraphics[width=0.5\textwidth,clip=]{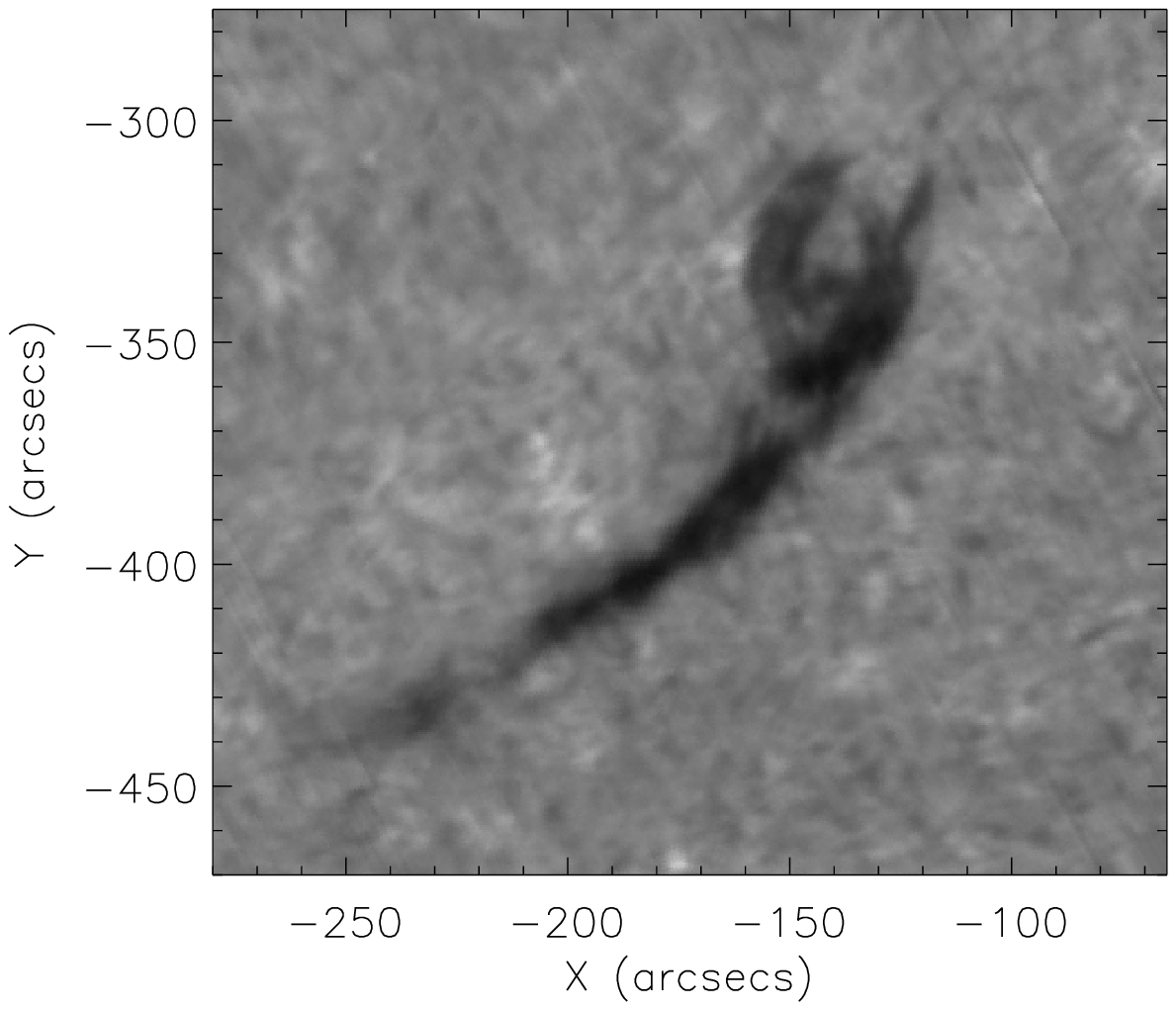}\includegraphics[width=0.5\textwidth,clip=]{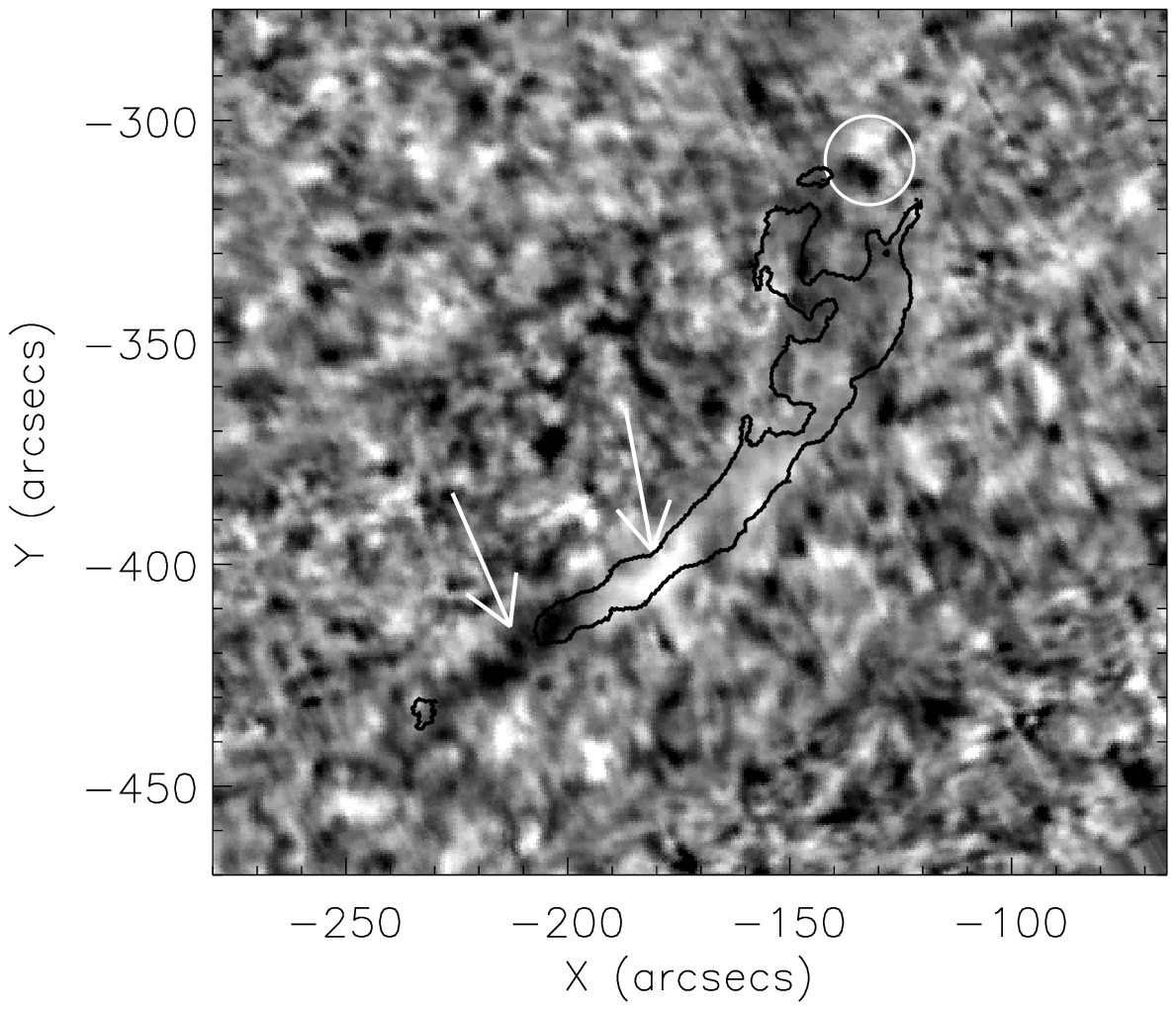}}
\caption{ The intensity and velocity images in  H$\alpha$ line observed using MSDP instrument at Meudon Solar Tower on 21 May 2008, at 09:00 UT. The left panel shows the line center intensity while the right panel shows the dopplergram overlaid by contours of the filament. The maximum (most white) and minimum (most black) velocities are +800 m s$^{-1}$ m s$^{-1}$m s$^{-1}$ and -200 m s$^{-1}$ respectively (positive is redshift). The white circle indicates up and downward flows at the end of one feet of the filament. Elongated areas of blue and redshifts suggest some twist along the filament body. }
\label{fig:msdp}
\end{figure}

\subsection{ H$\alpha$ images: Meudon Solar Tower and Udaipur Solar Observatory}

The Meudon Solar Tower observed the intensity and velocity field of
this region in H$\alpha$ line using MSDP instrument \cite{Mein77}
; \cite{Mein91}. The field of view was 295$\times$460 arcsec
with a pixel size of 0.5 arcsec. The morphological evolution of the
filament was covered by the H$\alpha$ filtergams recorded at Udaipur
Solar Observatory using Halle birefringent filter with 0.5 \AA\
bandpass filter centered at H$\alpha$. Our observing times are
summarized in Table 1. Figure~\ref{fig:uso} shows the evolution of the
morphology of the filament during the period 20-22 May. The filament
initially had thick spine oriented in North-South direction, later
became diffuse and patchy  and subsequently disappeared. The
disappearing phase was slow and lasted several ($\sim$ 10) hours.
 Figure~\ref{fig:msdp} (right) shows the H$\alpha$ dopplergram on 21 May at 09:00 UT. The two arrows shows the part of
the filament with blue-shifted (black) part corresponding to $\sim$ 200
m s$^{-1}$ and red-shifted (white) part corresponding to $\sim$ 800 m s$^{-1}$. These
elongated areas of blue and red-shifts suggest some twist along the
filament body \cite{Schmieder85b}. The contours mark the boundary of the filament which
is shown in  the right panel of Figure~\ref{fig:msdp}. We have made movies of
the MSDP H$\alpha$ dopplergrams and notice that the blue-shift is
always seen near the southern part of the filament suggesting that the
material is escaping from the lower end of the filament.
However, we do not see very strong velocities during the filament
disappearance in our H$\alpha$ observations, probably due to slow rise and
eruption of the filament. Also, the
small value of blue-shift of $\sim$ 200 m s$^{-1}$  is consistent with slow
disappearance of the filament.  However, these are line-of-sight velocities and as shown later in section 2.3.4, that the filament sheet is inclined to solar normal by about $\gamma=$ 47$^\circ$. Thus, applying $1/cos(\gamma$) correction to these velocities, we get blue-shifts of $\sim$ 300 m s$^{-1}$ and red-shifts of $\sim$ 1200 m s$^{-1}$.  The white circle in Figure~\ref{fig:msdp}
indicates up and downward flows at the end of one feet of the
filament.  Up and down flows at the feet are frequently observed  \cite{Schmieder91}, suggesting that filament could be formed by plasma injection through the barbs. Unfortunately, it has been difficult to prove that these motions like spicule motions are frequent or continuous enough to feed the prominence.  Higher resolution observations may prove it in the future.

\begin{figure}    
\centerline{\includegraphics[width=0.6\textwidth,clip=]{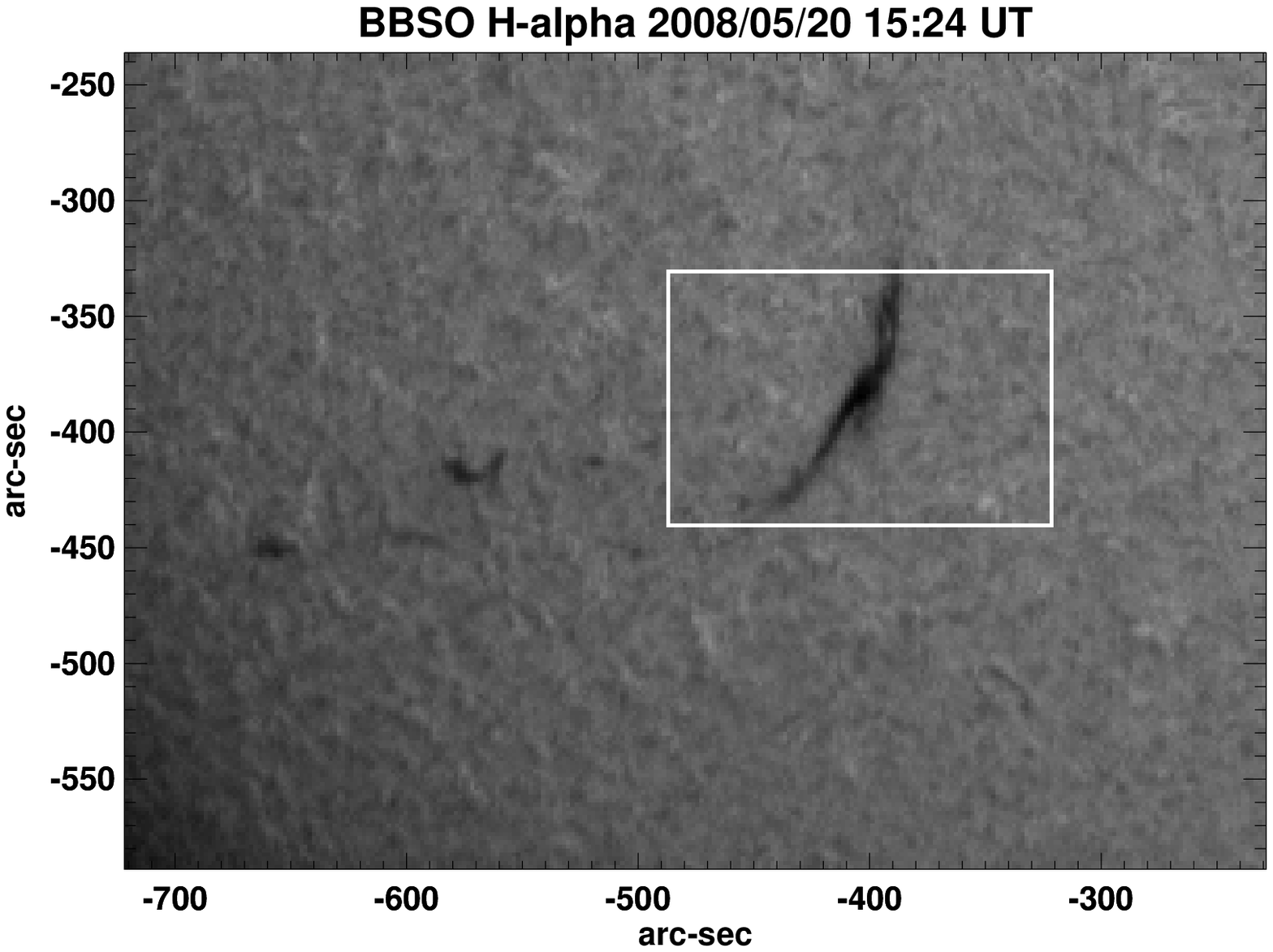}\includegraphics[width=0.6\textwidth,clip=]{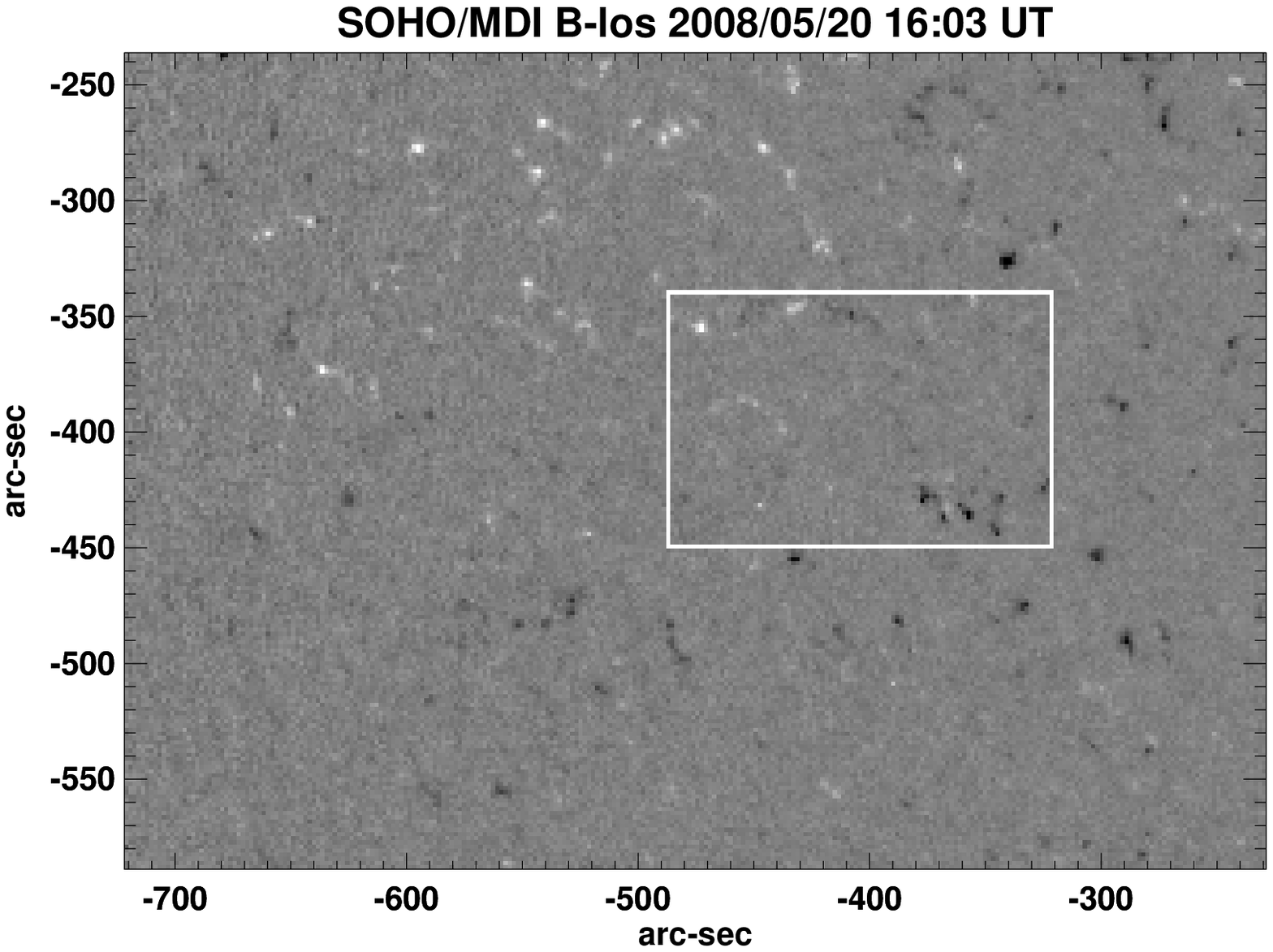}}
\centerline{\includegraphics[width=0.6\textwidth,clip=]{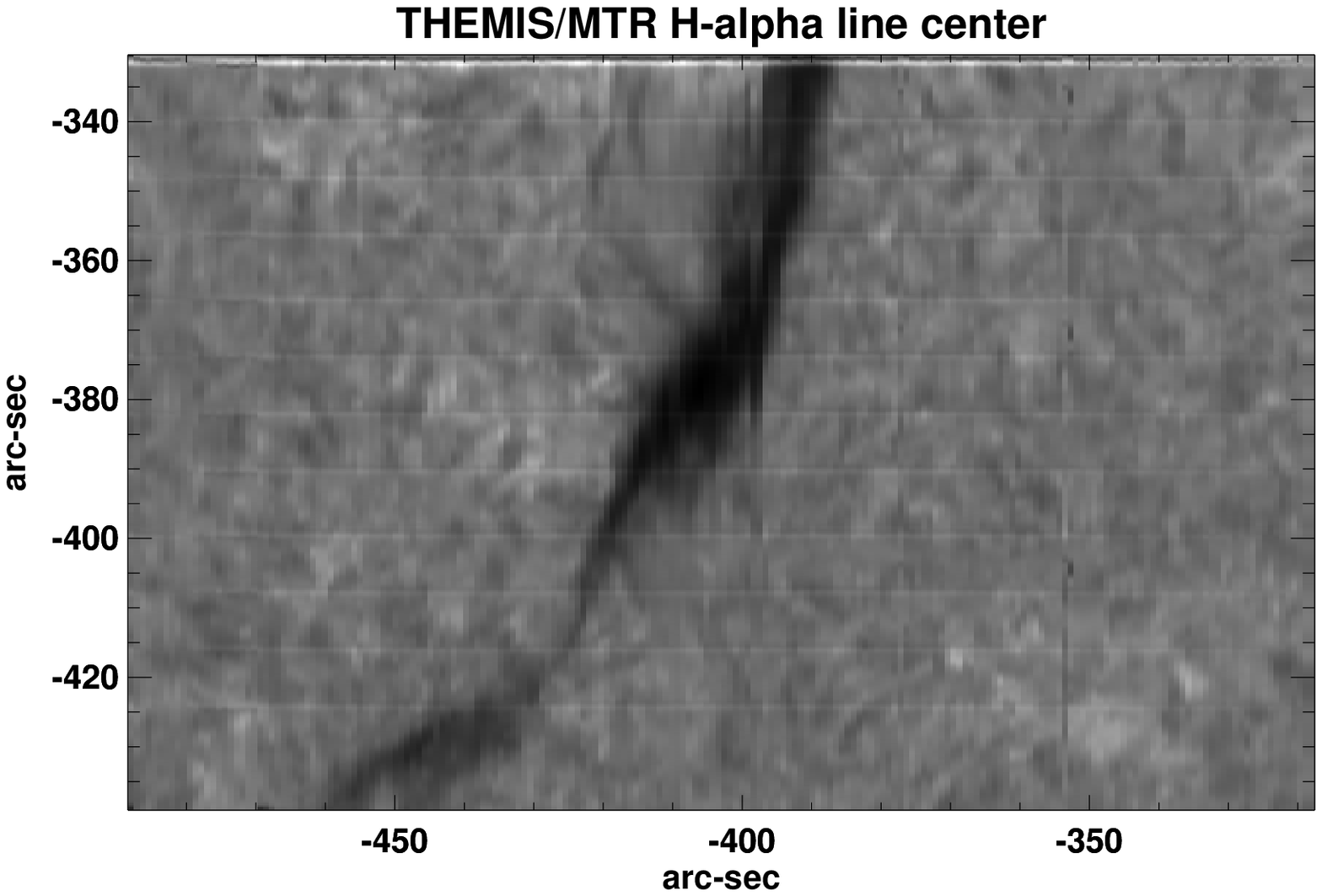}\includegraphics[width=0.6\textwidth,clip=]{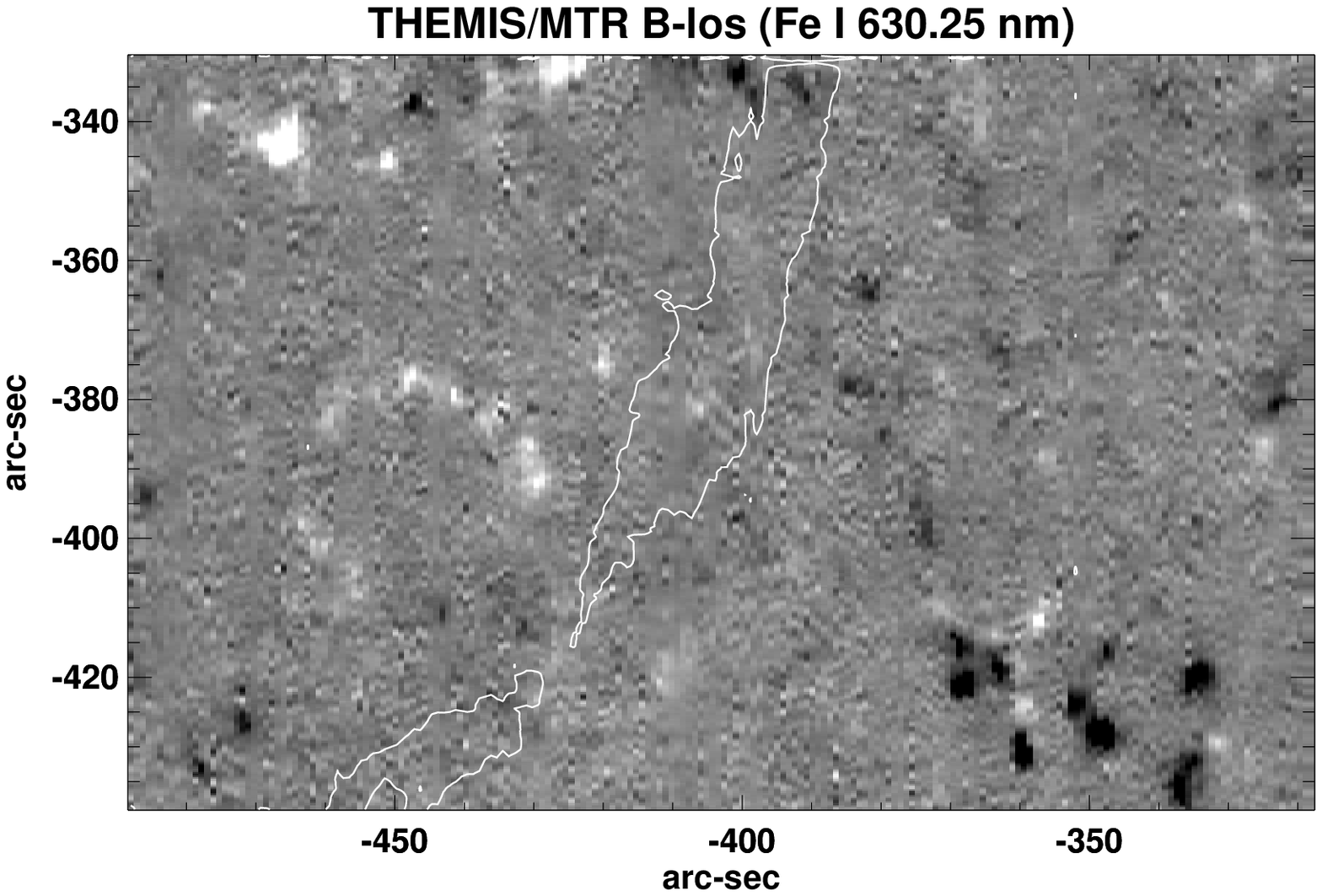}}
\centerline{\includegraphics[width=0.65\textwidth,clip=]{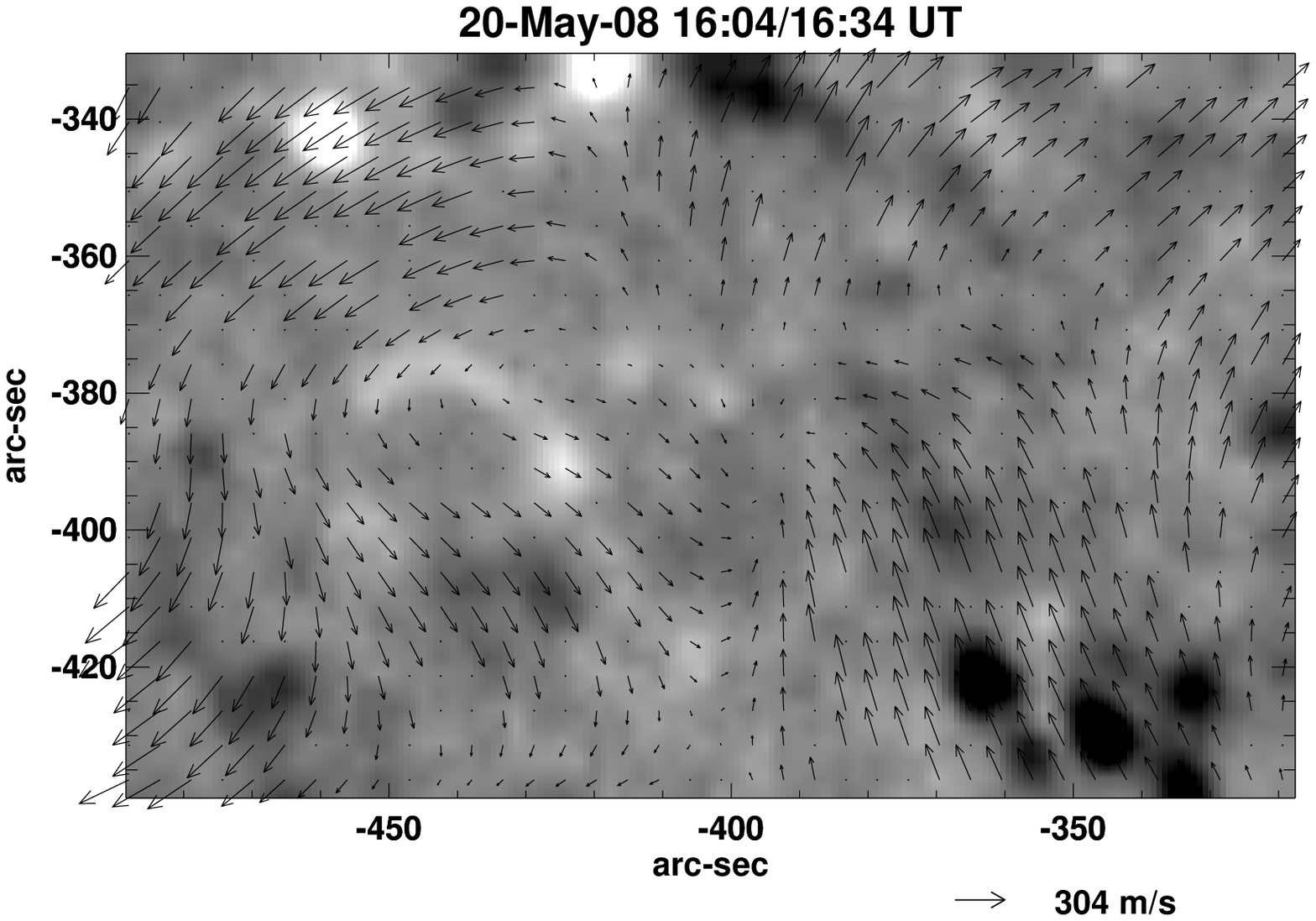}}
\caption{ {\it Top panel} shows the position of the filament  on the fulldisk H$\alpha$ image (left) and magnetogram (right) observed by  BBSO and SoHO/MDI respectively.  The date and time of observations are indicated on the top. The THEMIS/MTR FOV is indicated by a box on the fulldisk images.  {\it Middle panel }shows the THEMIS MTR scan of 20 May, (Seq. No. 14, 15:43 UT) with filament in H$\alpha$ (left) and Fe I 6302 \AA\ longitudinal flux (right). The longitudinal flux is scaled between +/- 250 Gauss in the full-disk magnetogram and between +/- 50 Gauss in THEMIS/MTR map. The filament is outlined by white contour. {\it Bottom panel } shows the horizontal flow velocity derived by applying Local Correlation Technique (LCT) to the region corresponding to MTR FOV, extracted from GONG full-disk magnetograms at the times indicated on top of the panel. }
\label{fig:mtrset1}
\end{figure}

\begin{figure}    
\centerline{\includegraphics[width=0.5\textwidth,trim=15 0 0 0]{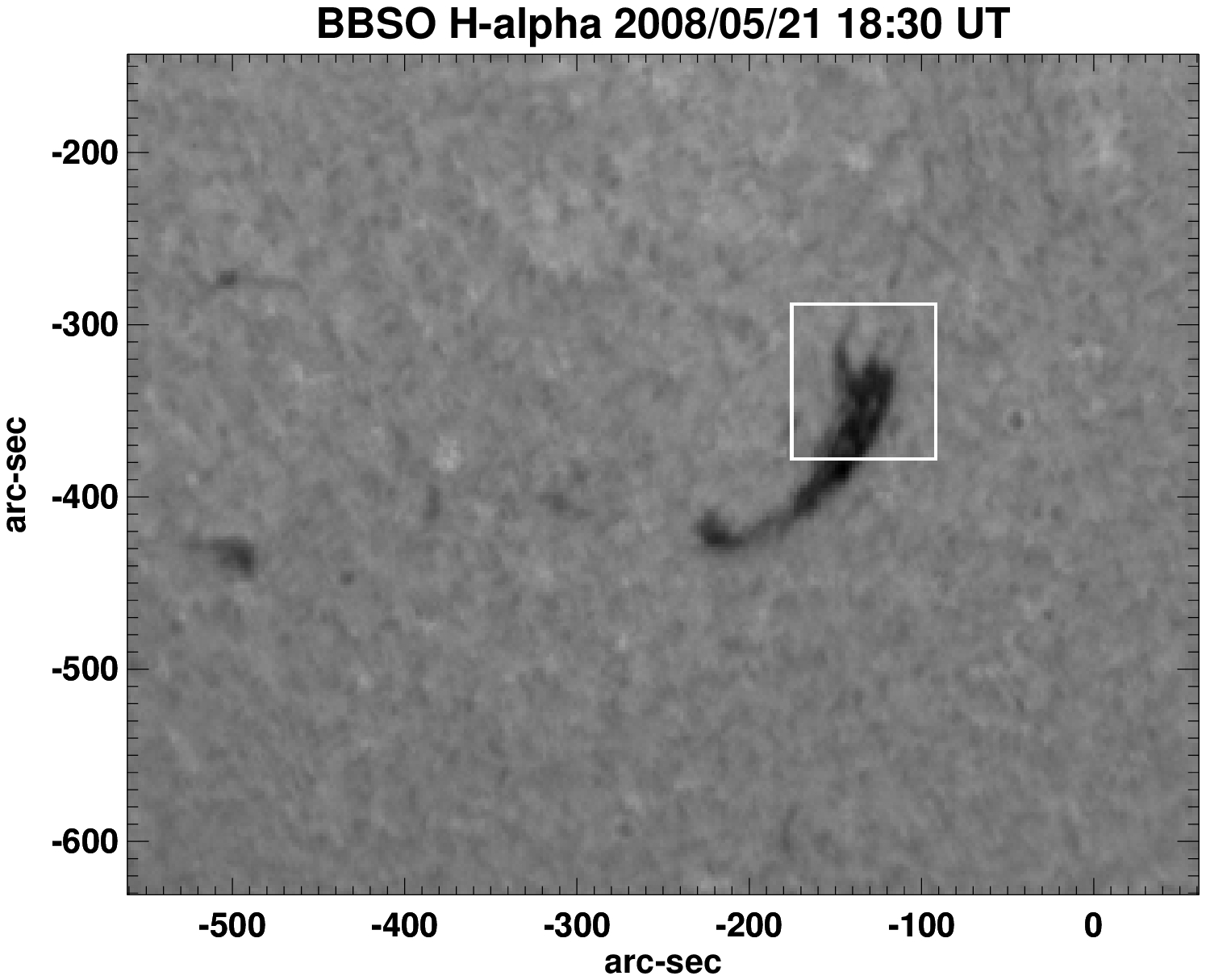}\hspace{0.15in}\includegraphics[width=0.5\textwidth,trim=15 0 0 0]{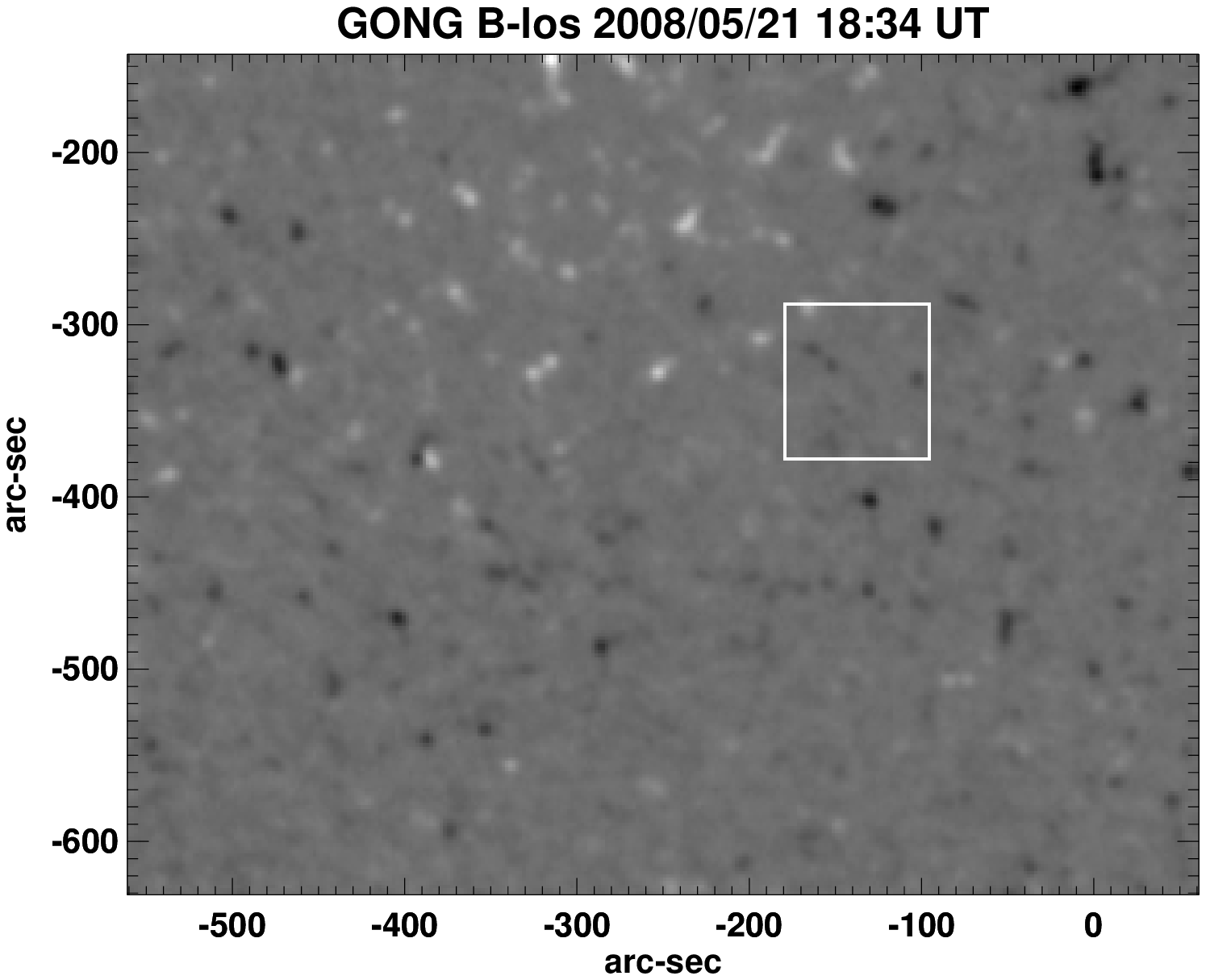}}
\vspace{0.1in}
\centerline{\includegraphics[width=0.35\textwidth,trim=0 0 0 0]{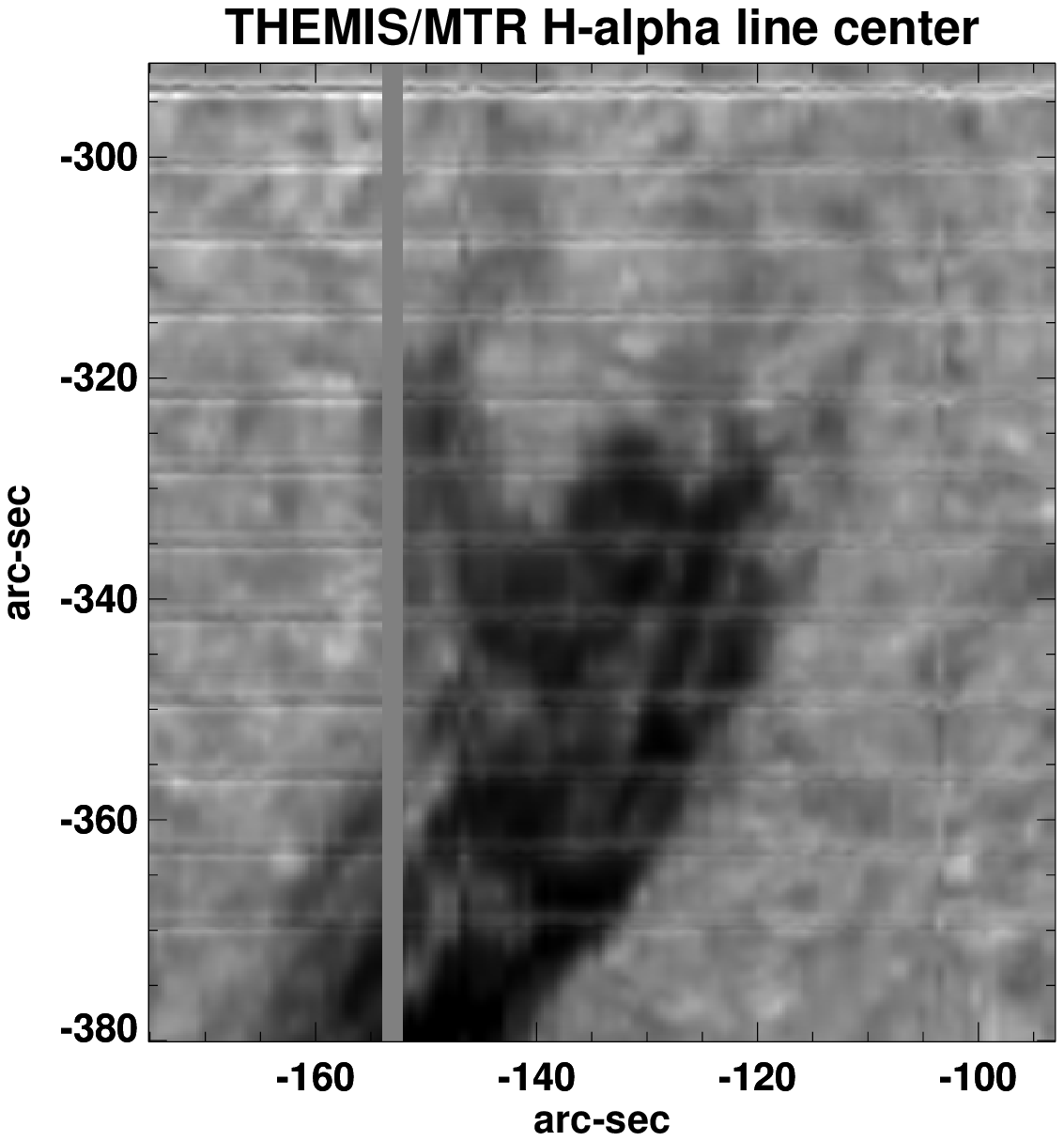}\includegraphics[width=0.35\textwidth,trim=0 0 0 0 ]{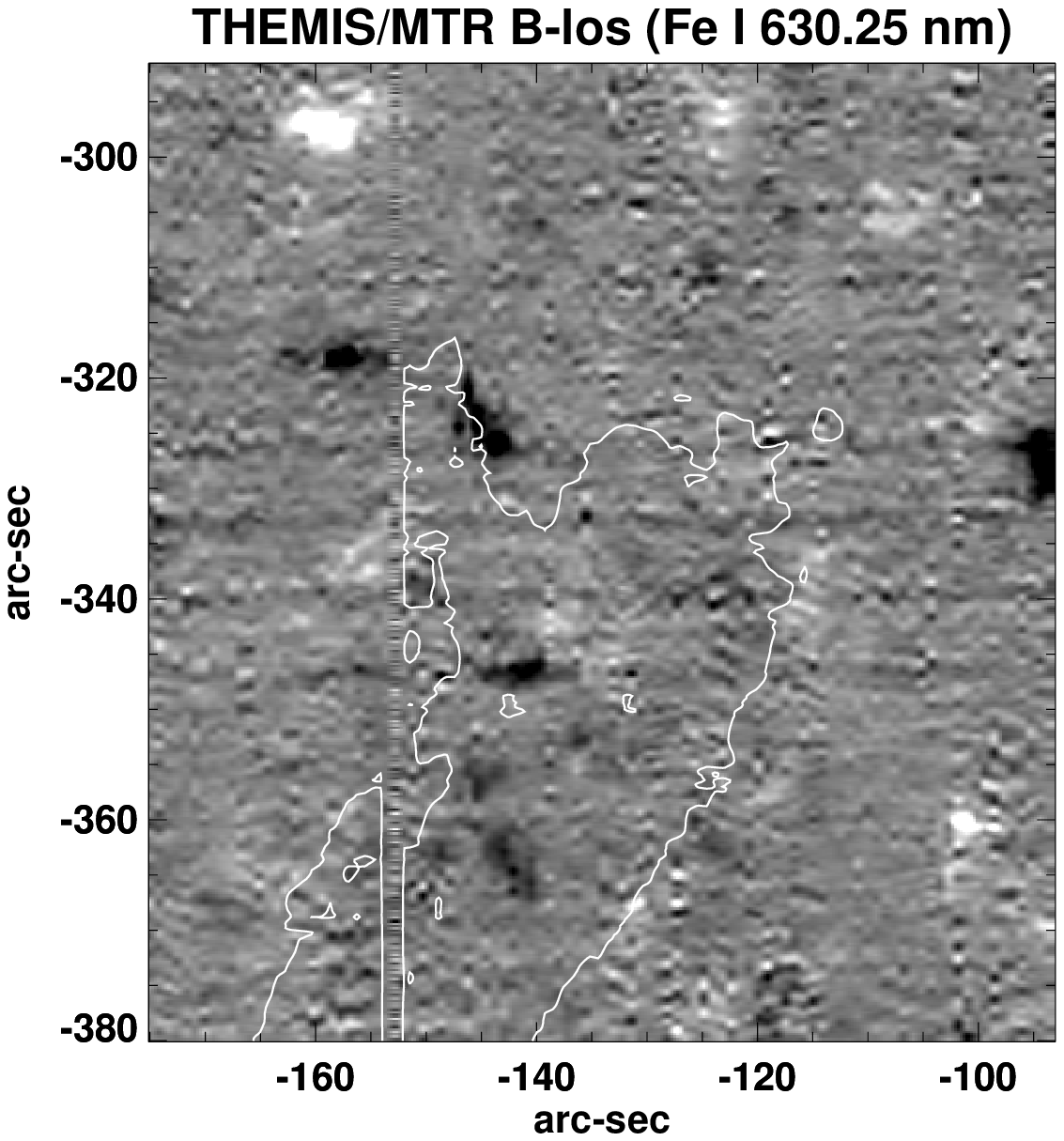}\includegraphics[width=0.35\textwidth,trim=0 15 10 0]{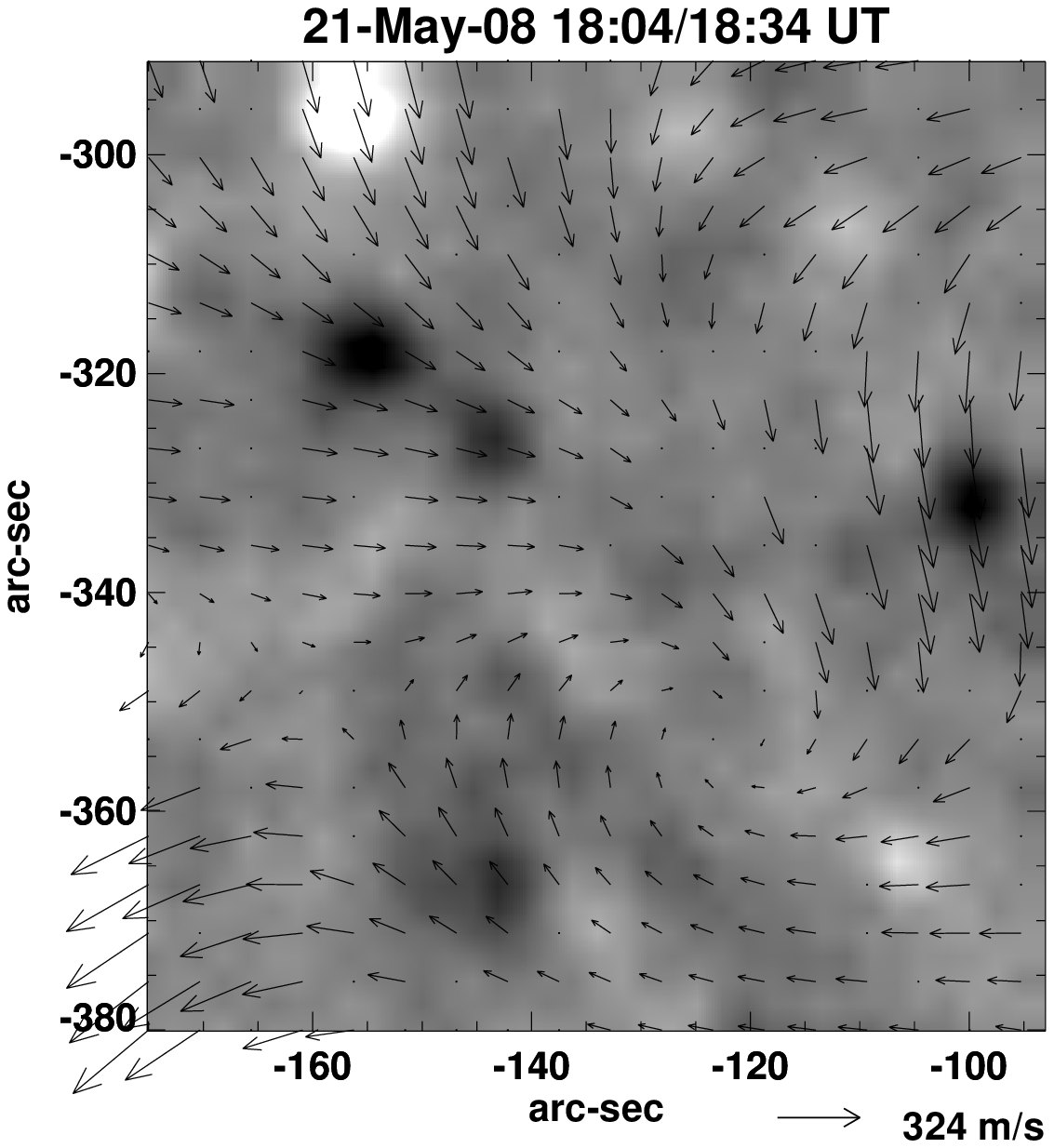}}
\caption{ Same as Figure~\ref{fig:mtrset1} but for 21 May , Seq. No. 16, 18:12 UT.}
\label{fig:mtrset2}
\end{figure}

\begin{figure}    
\centerline{\includegraphics[width=.498\textwidth,clip=]{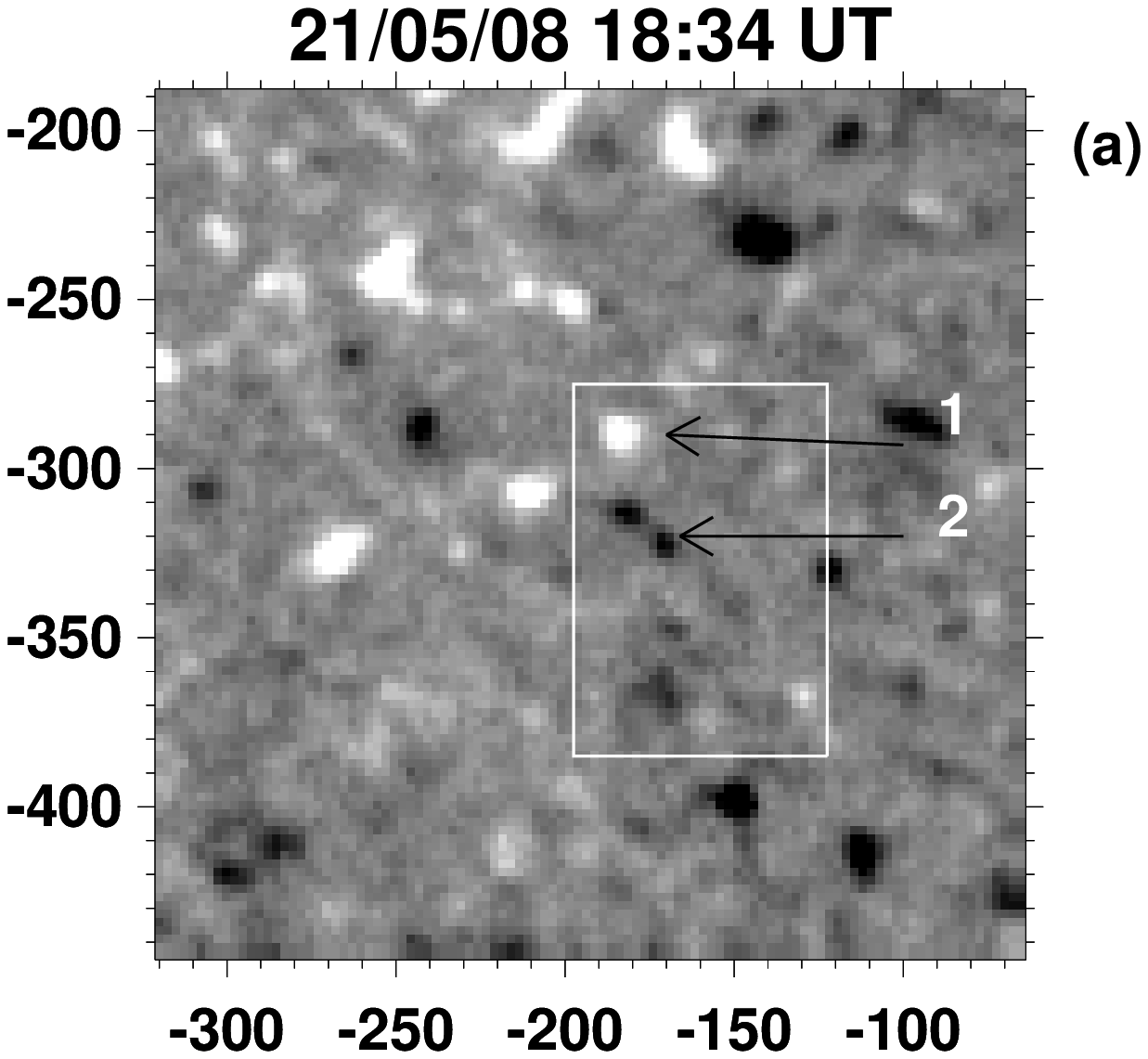}\includegraphics[width=0.45\textwidth,clip=]{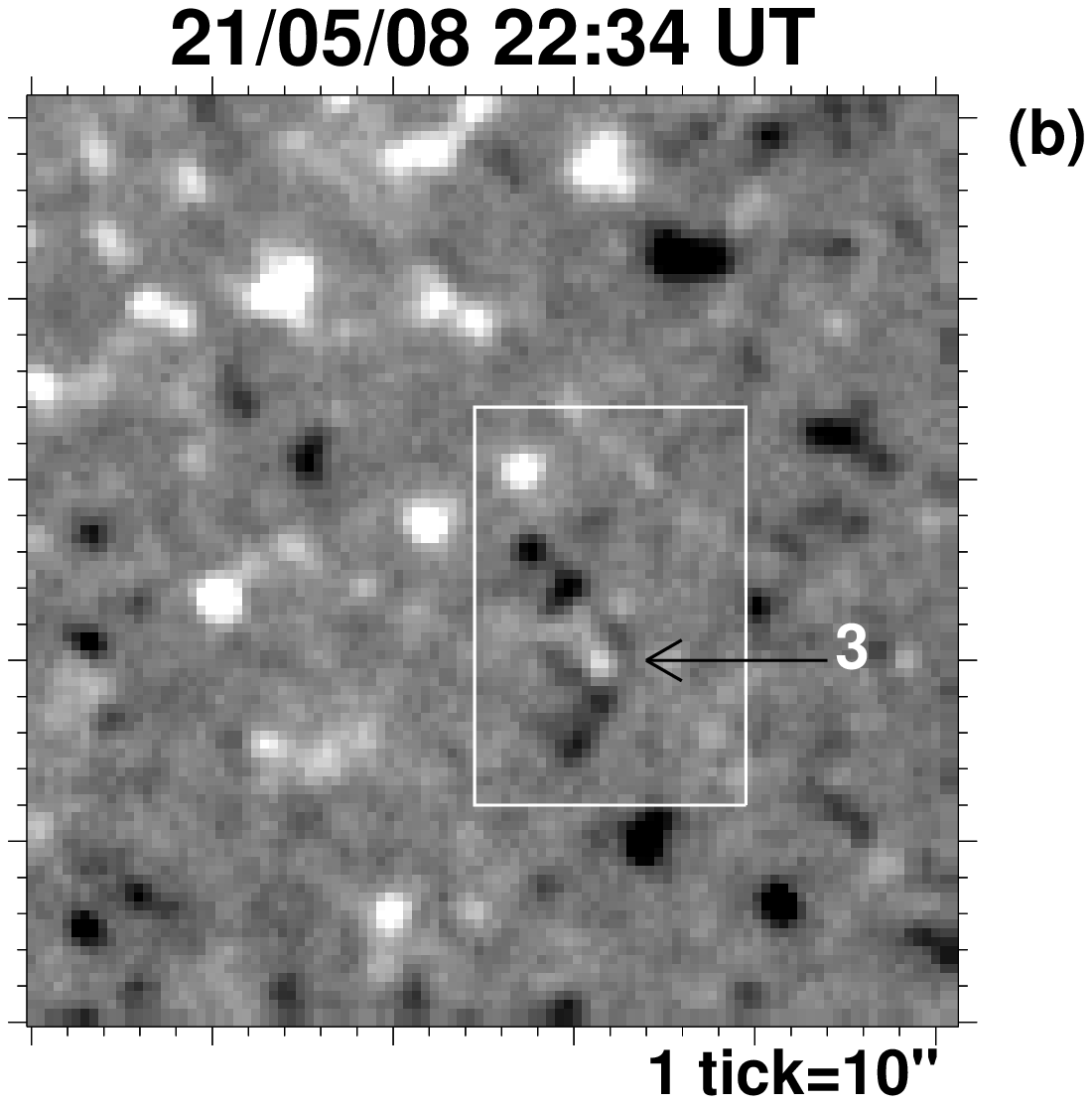}}
\vspace{0.1in}
\centerline{\hspace{0.2in}\includegraphics[width=.45\textwidth,clip=]{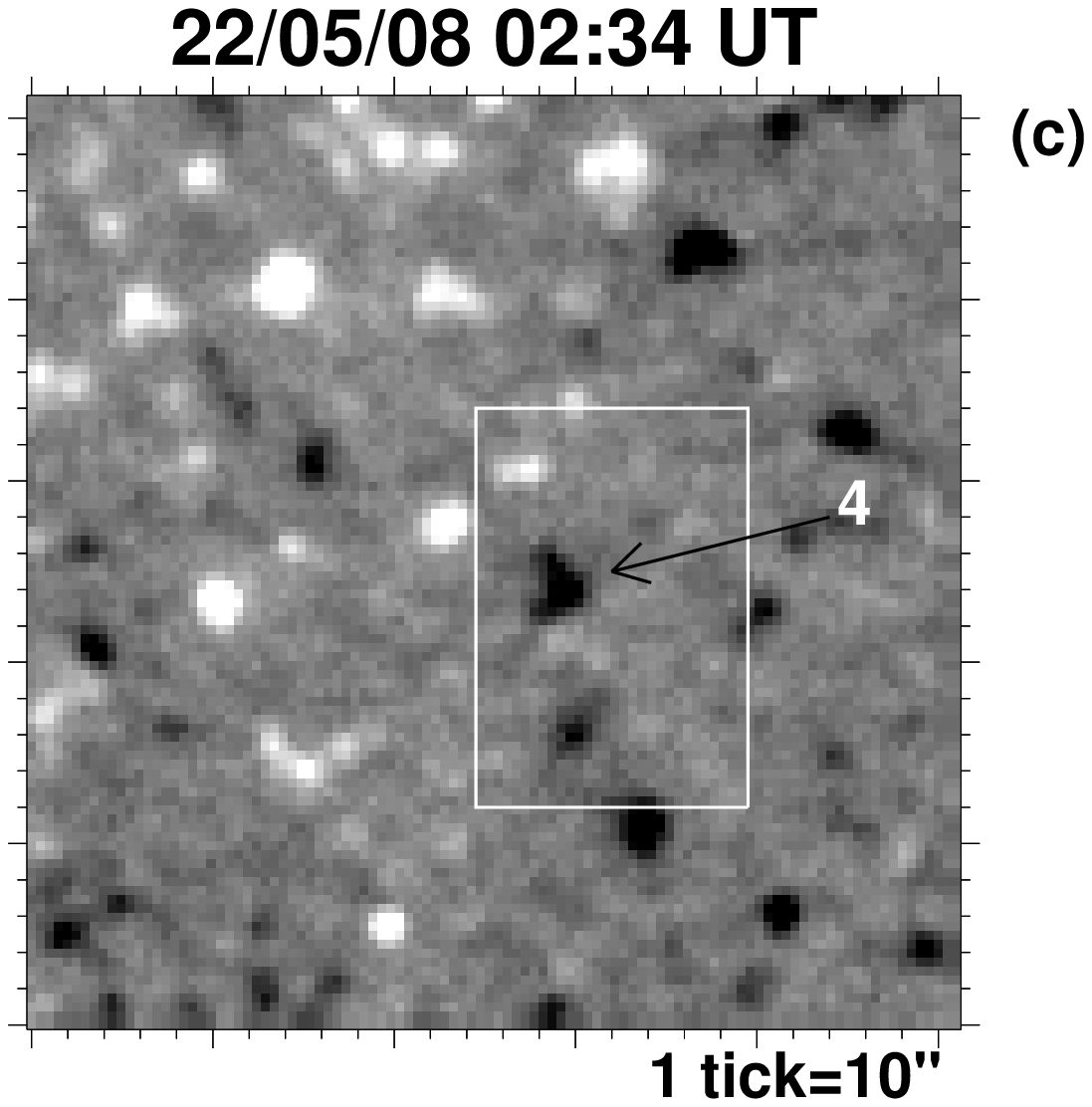}\includegraphics[width=0.45\textwidth,clip=]{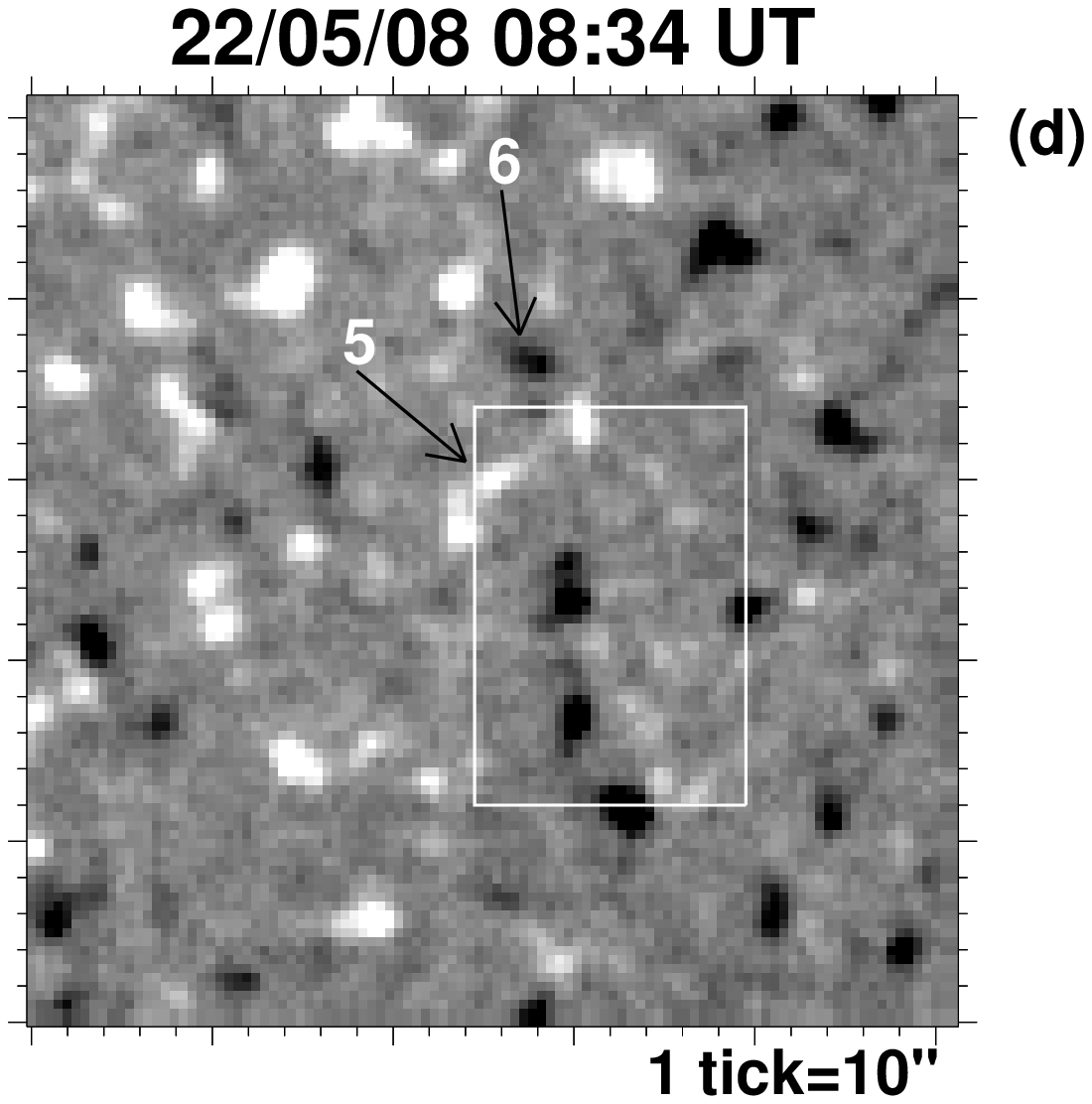}}
\caption{Evolution of the polarities near the filament feet (shown by arrows 1 and 2) which are recognized in MTR scan in Figure~\ref{fig:mtrset2} (shown here as white box). The new parasitic polarities are  indicated by arrows 3 and 6 . The polarities 2 and 1 have evolved into polarities marked by arrows 4 and 5 respectively. }
\label{fig:gongpolarity}
\end{figure}

\begin{figure}    
\centerline{\includegraphics[width=0.5\textwidth,trim=10 0 0 0 ]{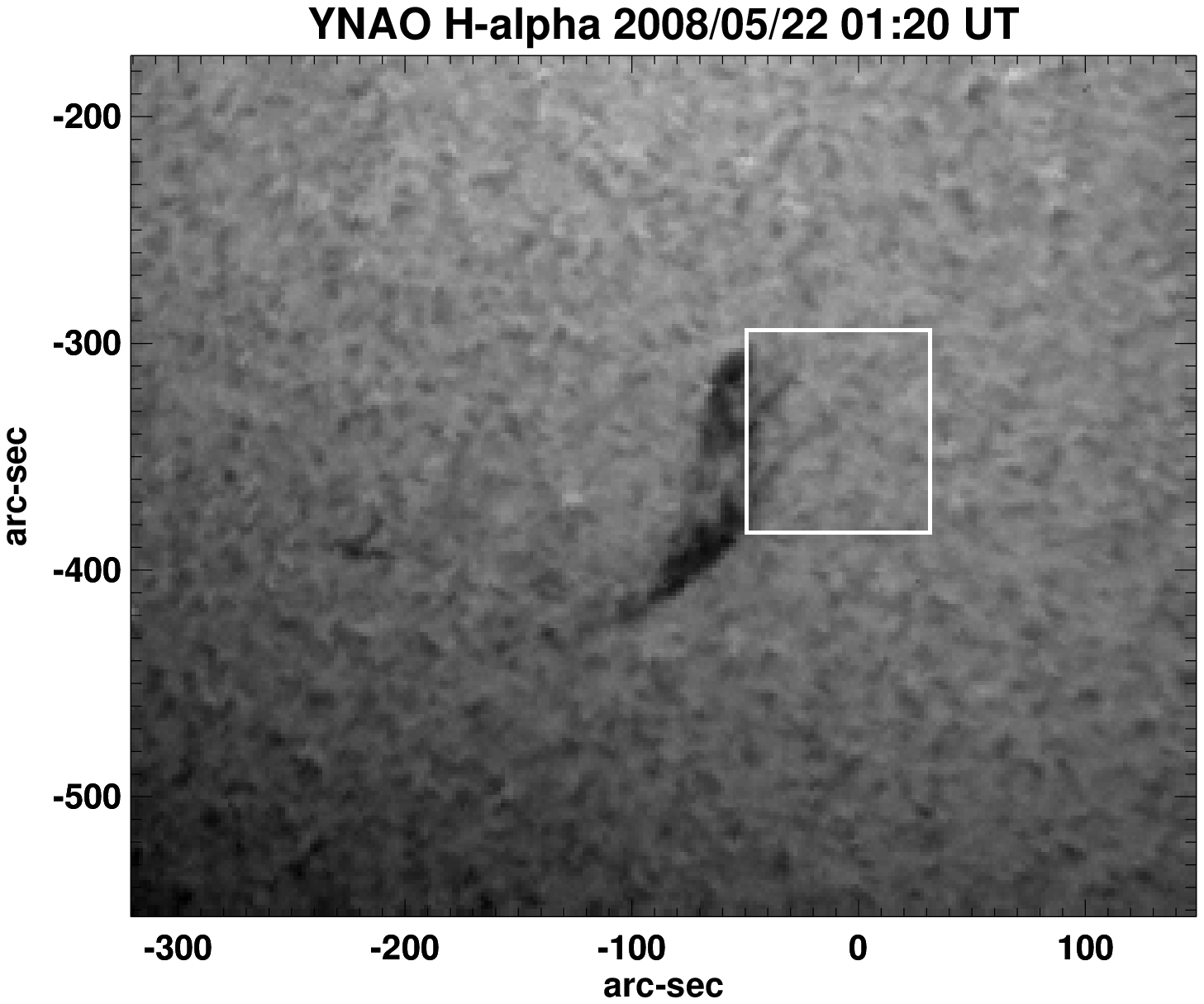}\hspace{0.15in}\includegraphics[width=0.5\textwidth,trim=10 0 0 0]{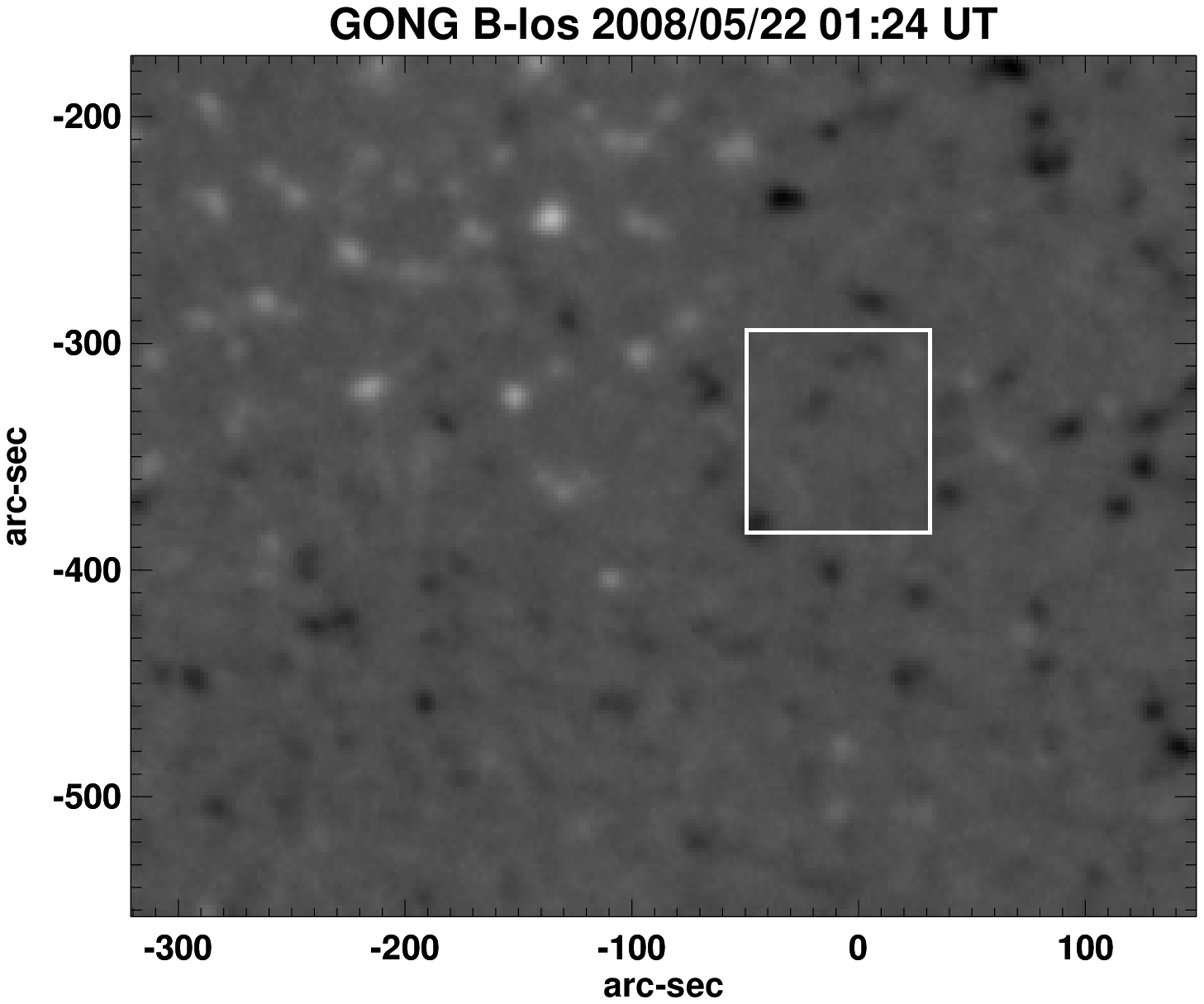}}
\vspace{0.15in}
\centerline{\includegraphics[width=0.31\textwidth,trim=15 15 0 0]{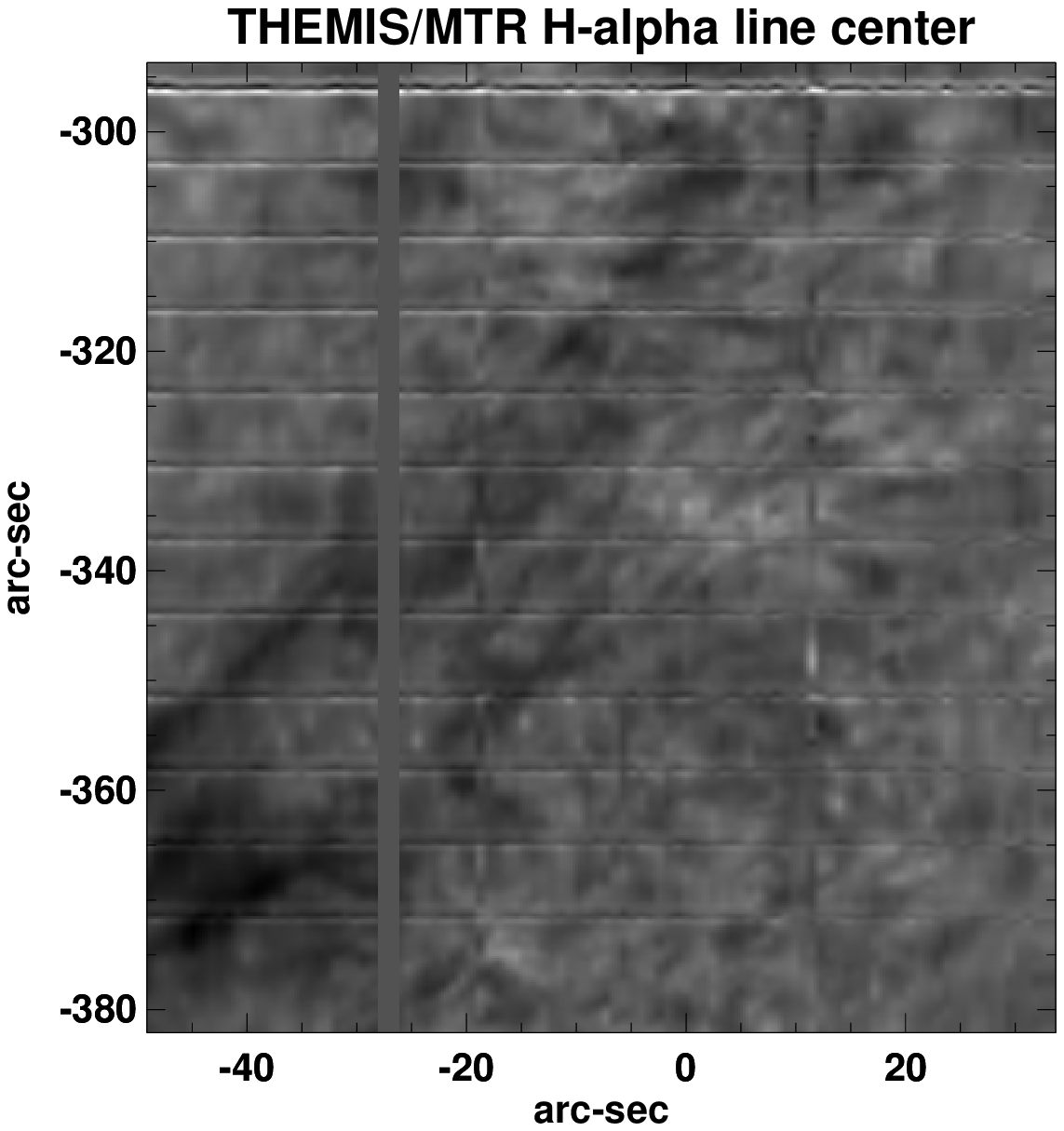}\hspace{0.1in}\includegraphics[width=0.32\textwidth,trim=5 15 0 0]{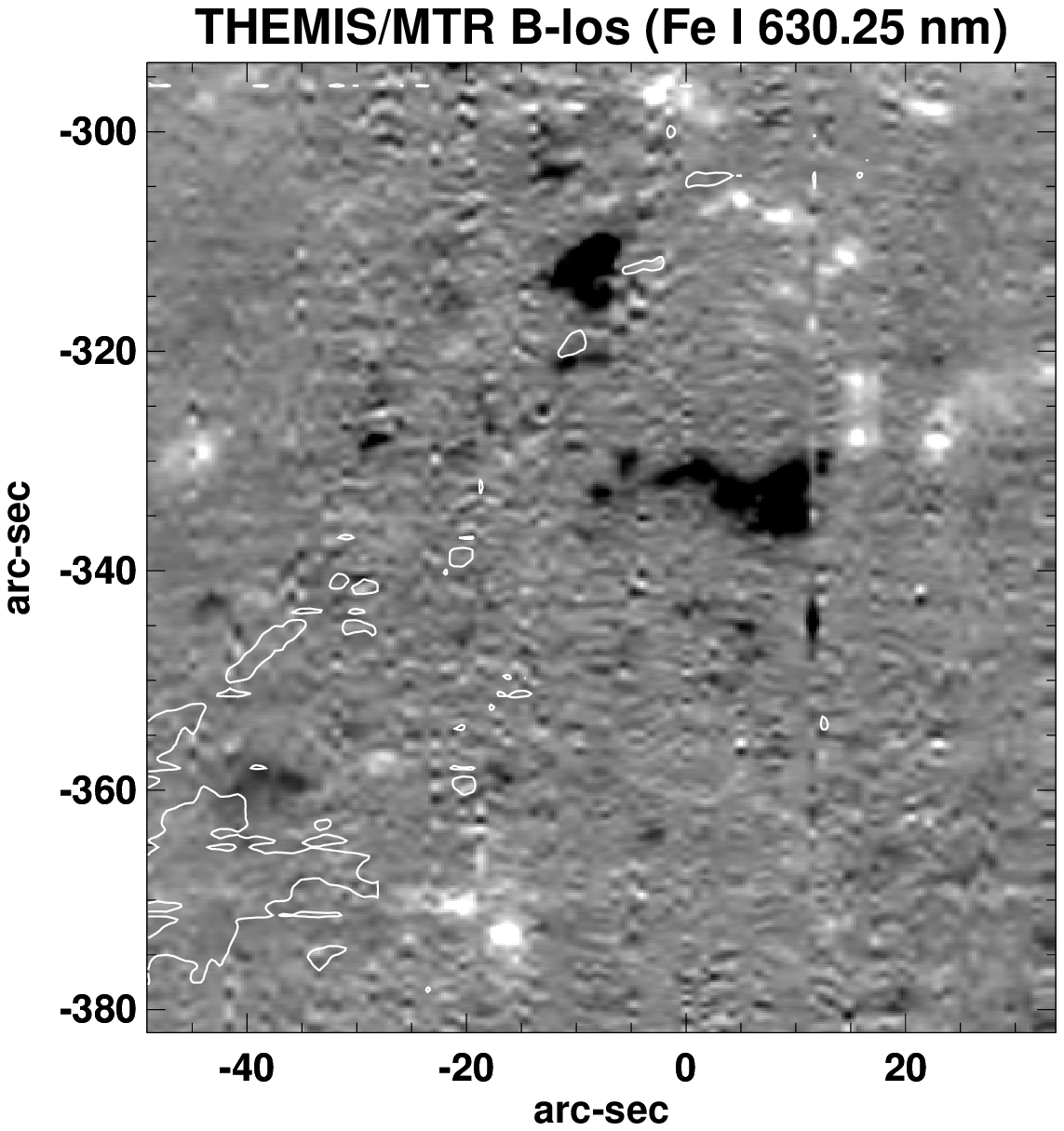}\hspace{0.1in}\includegraphics[width=0.32\textwidth,trim=15 25 0 0]{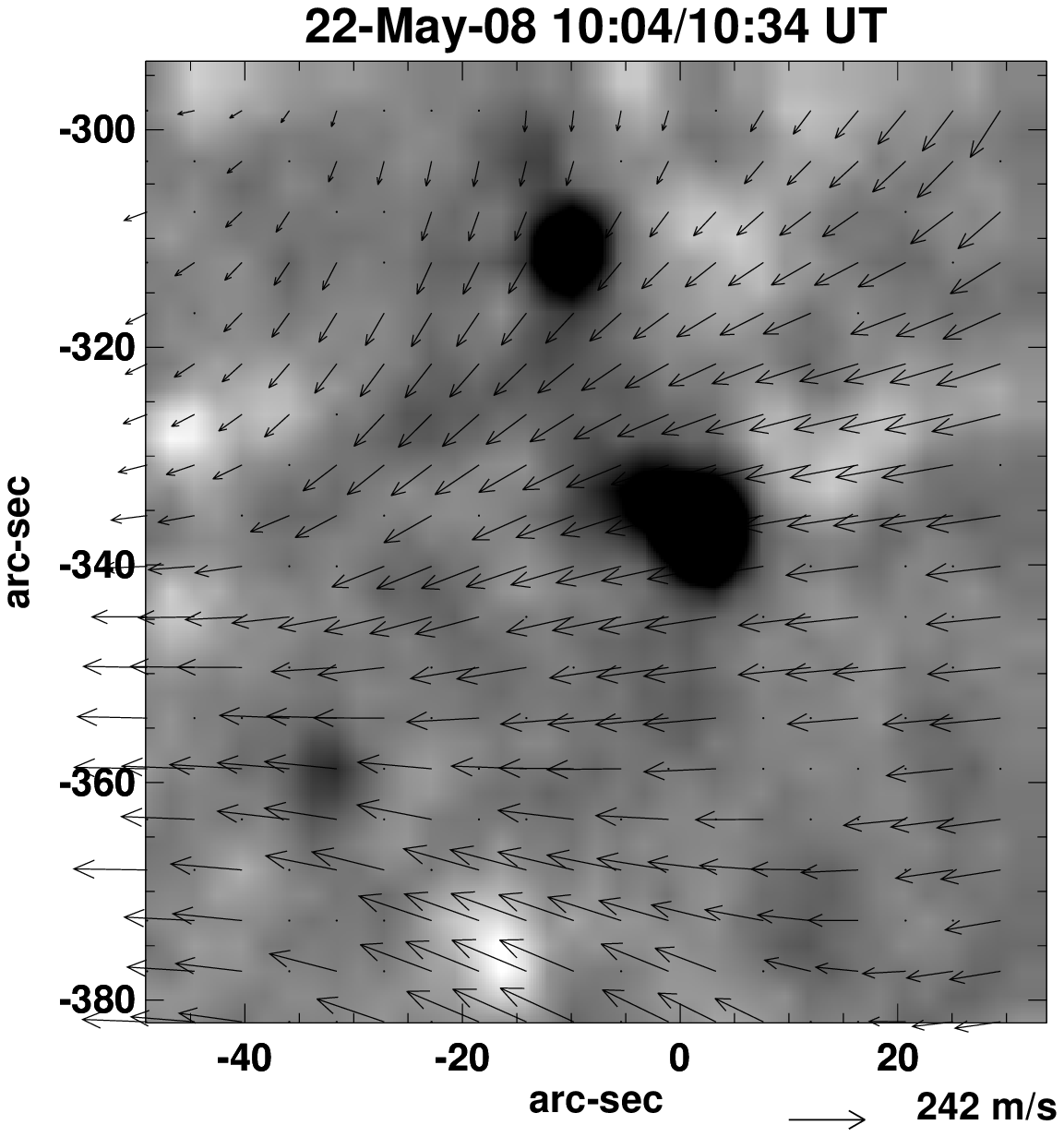}}
\vspace{0.15in}
\caption{ Same as Figure~\ref{fig:mtrset1} but for 22 May , Seq. No. 2, 09:47 UT.}
\label{fig:mtrset3}
\end{figure}

\subsection{ THEMIS-MTR and GONG observations}

 The high resolution spectro-polarimetric observations of the filament were carried out using MTR instrument of THEMIS  \cite{2002A&A...381..227B},\cite{2005A&A...435.1115B},\cite{Guo09}.
  We choose the following three sets from MTR for this analysis (1) 20 May: seq. 14, during the early stable stages of the filament. This gives the contextual information, (ii) 21 May: seq. 16, the filament has evolved into a more diffuse patchy and broad configuration, and (iii) 22 May: seq. 2, the filament is in disappearing phase. The polarities near the feet of the filament which is detaching from the solar surface are studied. The summary of these three sets are shown in Figure~\ref{fig:mtrset1},~\ref{fig:mtrset2} and~\ref{fig:mtrset3} respectively, with top panels showing full-disk H$\alpha$ images and longitudinal magnetograms. The THEMIS/MTR field-of-view (FOV) is indicated by a box, while middle panels show the H$\alpha$  and longitudinal magnetic field map derived from THEMIS/MTR scans. The MTR observations of the filament channel were taken in Fe I 6302 \AA\, H$\alpha$, Na 5896 \AA\ and Fe 5250 \AA\ lines. The photospheric magnetic field was derived by fitting Fe I 6302 \AA\ lines with the profiles computed using UNNOFIT \cite{Bommier07}. The transverse magnetic field is too weak and hence is not shown in these inverted maps. The pattern of weak magnetic polarities in the filament channel as seen by high sensitivity of THEMIS/MTR measurements is recognized in full-disk GONG/MDI magnetograms as close in time as possible. Since the evolution of magnetic field using spectral scans is not feasible as scans take long time for a single magnetic map, we use full-disk longitudinal  magnetograms from GONG instrument to study time evolution of magnetic field in the filament channel.

   The evolution of magnetic polarities using GONG magnetogram is studied in these steps:

  (1) MTR scans in H$\alpha$ and Fe I 6302 \AA\ line are automatically co-aligned since these are strictly simultaneous.  So, the magnetic polarities of interest (POI), like polarities near filament feet or barbs or parasitic polarities are conveniently located using overlays of MTR H$\alpha$ over MTR Fe I 6302 \AA\ maps.

 (2) These POI are then identified in fulldisk GONG magnetogram by matching the pattern visually. A good correspondence between GONG and MTR magnetogram panels in Figure~\ref{fig:mtrset1}, 6 and 8 is evident. The appropriate region is then extracted from GONG full-disk magnetograms and co-aligned using cross-correlation technique.

Thus, the co-alignment of ground-based datasets is required only among successive GONG magnetograms for Local Correlation Tracking (LCT) analysis and for making  movies. GONG magnetograms are chosen because these are available most of the time in a network of six observing stations. Also, during our campaign the MDI magnetograms are available with large data gaps and therefore not used much in this work.

\subsubsection{ Evolution of polarities near filament barbs}
  The feet or the barbs are visible in THEMIS/MTR H$\alpha$ image shown in Figure~\ref{fig:mtrset2}. Figure~\ref{fig:gongpolarity} shows the evolution of the polarities near the feet of the filament.  These polarities are shown in panel ~\ref{fig:gongpolarity}(a) with arrows 1 and 2. These were identified using THEMIS/MTR scan as shown in Figure~\ref{fig:mtrset2}. This scan is represented by white box in Figure~\ref{fig:gongpolarity}. In the panel ~\ref{fig:gongpolarity}(b) the arrow 3 shows new parasitic polarity (positive) emerging near point 2. In panel  ~\ref{fig:gongpolarity}(c) this parasitic polarity has disappeared. Further, the two negative polarity patches corresponding to arrow 2 coalesce to form the patch shown by arrow 4. Similarly, in panel ~\ref{fig:gongpolarity}(d) the arrow 5 corresponds to polarity 1,  merging with other positive polarity patches. The arrow 6 corresponds to another new parasitic polarity (negative) that has emerged in the channel. To visualize the evolution of polarities in the filament channel in general, we made a movie of the LOS magnetic field with these co-aligned GONG magnetograms. These movies are made for the duration  20 May, 09:04 UT to 22 May, 22:34 UT, which covers the entire period of THEMIS observations and also the filament disappearance event. These movies are made available online in the electronic version of this paper [see, movie-1 on {\it http://tinyurl.com/movie-html}]. The polarities in THEMIS/MTR FOV as indicated in Figure~\ref{fig:gongpolarity} by the white box, can be recognized in these movies. The magnetic field in the channel is weak, about $\pm$ 10 to $\pm$40 Gauss, and changes over tens of hours. The  movie shows parasitic polarities emerging and disappearing in and around the filament channel continuously. These changes can induce the change of the footpoint barb  and the  untwisting flux rope \cite{2006SoPh..238..245S}.

\subsubsection{ Horizontal flows : Local Correlation Tracking}

 Further, the plasma flows in the filament channel are  studied by computing horizontal flow velocities using local correlation technique (LCT) \cite{Welsch04}, \cite{2008A&A...480..255R}. We use GONG magnetograms, separated half-an-hour apart, for computing these LCT maps during the entire period 20, 21 and 22 May. The flow maps for 20, 21 and 22 May are shown in the lower panels of Figure~\ref{fig:mtrset1}, ~\ref{fig:mtrset2} and ~\ref{fig:mtrset3} respectively, corresponding to the FOV and observing times of THEMIS/MTR.  The flow velocity derived by LCT during the observations is of the order of 250 to 350 m s$^{-1}$. The LCT map for 20 May shows the shearing flow pattern, with flow vectors pointing upwards in the right side and downwards on the left side of the map, while the filament is located in between these flows. The flow field is changing continuously and on 21 May, the pattern is like a vortex. For the location of filament see white contour for 21 May THEMIS/MTR FOV. During 22 May, the flow pattern is more or less unidirectional in this small FOV and is oriented towards the remnant filament seen in 22 May THEMIS/MTR FOV.  The increase of the shear flow leads to an increase in magnetic shear and destabilizes the filament. The presence of vortex flow favours canceling of magnetic flux at the location of filament feet. These motions could be responsible for canceling of flux, and later on disappearance of the foot-point barbs where the filament is tied to the photosphere and to the eruptions (Raadu {\it et al.} 1987,1988).

\subsection{STEREO SECCHI/EUVI He II 304 \AA\ observations}

STEREO/SECCHI/EUVI A and B spacecraft continuously observe the Sun in EUVI lines 171 (Fe IX), 195 (Fe XII), 284 (Fe XV) and 304 \AA\ (He II) from two different angles. These images give stereoscopic view of the extended objects like coronal loops and filaments.
The angle of separation between STEREO A and B during our observations is around 52.4 degrees.  In the following we present analysis of the stereoscopic He II 304 \AA\ observations of the filament using different methods.

\subsubsection{I. Analysis using movies}

We use stereoscopic observations from STEREO/SECCHI/EUVI A and B satellite to study evolution of filament
dynamics before and during its eruption.
We made movies of the event in both STEREO-A and B observations using He II 304 \AA\ filtergrams which were observed at a cadence of 10 minutes during the campaign. The movie is available in electronic version [see, movie-2 on {\it http://tinyurl.com/movie-html}].
The visual analysis of these observations in He II 304 \AA\ of the filament, clearly shows that unlike H$\alpha$, the  filament is extended
over very large distance. We notice that prior to its disappearance  the filament splits up into several thin filamentary structures parallel
 to filament axis, which subsequently disappear. The chirality of the filament appears to be sinistral as inferred by  sense of twist
 of the filament threads during the eruption. The chirality  inferred
 from the H$\alpha$ data using the direction of barbs \cite{2006A&A...456..725L} is also sinistral. Also, globally the  erupting filament
seen in He II 304 \AA\ filtergrams had  a long S-shaped PIL. Thus,
the filament follows the hemispheric chirality rules for filaments \cite{2003ApJ...595..500P}. Also
from the movies it seems that the filament evolves in two phases,
(i) the southern part evolves on 22 May around 2-3 UT, a faint CME
reported by LASCO C2 at 02:00 UT could be associated with this
part of the filament, and (ii) the north and west part which evolves
 between   10-12 UT.  LASCO was not observing around these times. STEREO COR1 and COR2 were observing and did not detect any CME, which implies either the CME was too faint to be detected or there was no CME at all.

\begin{figure}    
\centerline{\includegraphics[width=1.0\textwidth,clip=]{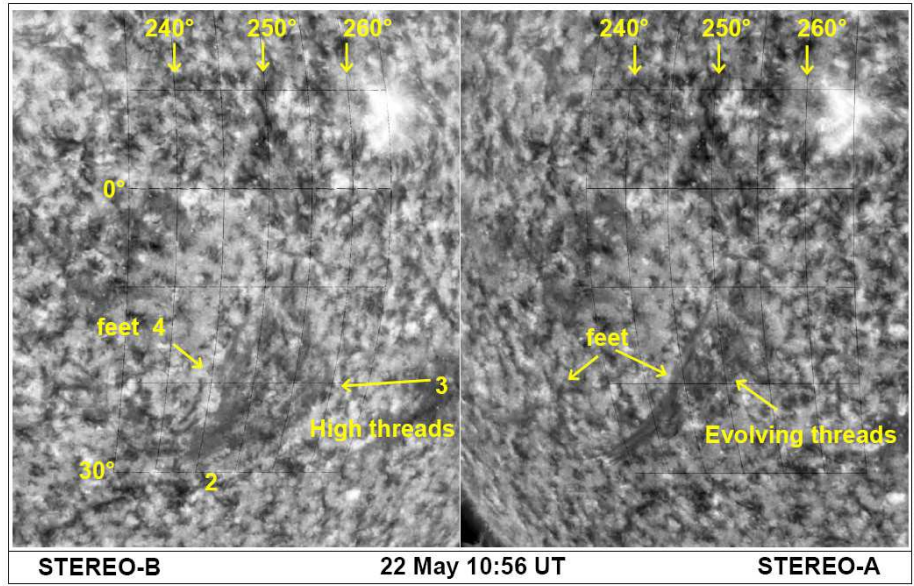}}
\caption{ The filament as seen in STEREO/EUVI A (right image) and B (left image) on 22 May 2008 at 10:56 UT.  Spherical  coordinates on a surface of 700 Mm have been drawn for
given longitudes and latitudes between 240 and 260 degrees and 0 to 30 degrees
respectively. Numbers 1 and 4 indicate the feet of the filament, number 2 and 3
the top threads.  Notice the fan shape of the filament.}
\label{fig:carrgrid}
\end{figure}

Further, from the movie we notice that the evolving threads observed by STEREO A, have  a fan shape and an oscillating nature before disappearing.
 Before escaping, a few threads (Longitude=$260^\circ$, Latitude=$20^\circ$, in right panel of Figure~\ref{fig:carrgrid}), oscillate three times,
  between 09:56 and 10:16 UT, between 10:46 and 11:06, and between 11:36 and 11:46 UT.  Large amplitude oscillations in a filament
  before disappearance have been reported earlier by \inlinecite{2007SoPh..246...89I} and \inlinecite{2008A&A...484..487C}.
 Further, the topmost threads seem to be rising with a projected velocity of about 100 km s$^{-1}$ with respect to the filament feet  as estimated from the movie.

\subsubsection{II. Analysis using spherical grid method}
Images of He II 304 \AA\ by STEREO/EUVI B (left image) and STEREO/EUVI A (right image) on 22 May, around 10:56 UT are shown in Figure~\ref{fig:carrgrid}.  Using the grid one can differentiate the surface features from elevated structures as they do not appear at the same latitude and longitude due to projection effects. In Figure~\ref{fig:carrgrid}, the numbers 1 and 4 indicate the feet of the filament, number 2 and 3 the high-lying threads of the spine.   There are clear indications that the filament sheet is not normal to solar surface,  but inclined. The following observations support this inference:\\
(i) Even though the longitude separation between the filament and central meridian (say $\phi$)  in frame of STEREO B ($\phi_B$=17 degrees)  is smaller compared to that of STEREO-A ($\phi_A$=35 degrees), the width  of the filament is  broader in STEREO B as compared to STEREO A.  Whereas, we know that for a filament sheet which is normal to solar surface, the observed width should increase with $\phi$ and reach maximum at the limb, where it would be seen as prominence. \\
(ii) If the filament sheet is normal to solar surface then the  filament sheet in STEREO-A should be seen projected on the right side of the filament feet. While, we can observe feet 4 in both STEREO A and B images on the same side  (on the left of the filament).\\

Thus, we infer that the filament sheet is inclined. Further, we can use the fact that filament sheet is projected on the same side (towards right) of the filament feet in both STEREO A and B, to put an upper limit on the filament inclination. We know that if the filament sheet is inclined  to the line-of-sight then the inclination is equal to angle $\phi$. In which case the filament feet will be difficult to observe as these will be obscured by filament itself. Thus, from observation (ii) above
we can say that filament inclination is greater than angle $\phi_A$, that is, 35 degrees. Indeed in section 2.3.4 we measure filament inclination using triangulation technique to be about 47 degrees.

\subsubsection{Height measurement using SCC\_MEASURE}
\begin{figure}    
\centerline{\includegraphics[width=1.0\textwidth,clip=]{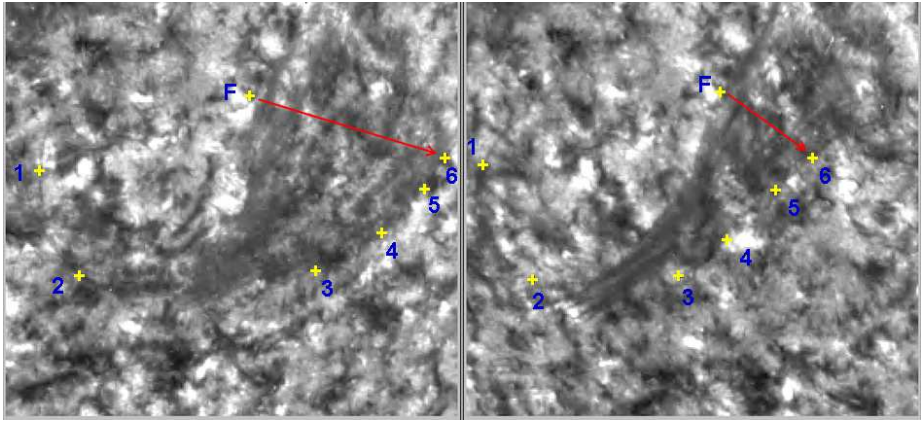}}
\caption{ The filament as viewed in STEREO-B (left panel) and STEREO-A (right panel) and the reconstruction of height along different points on the spine. The height in Mm from solar surface is given for locations marked by yellow '+' marks. The location numbers are marked in blue colors. The height determined from SCC\_MEASURE routine for these locations are mentioned in Table 2. Location marked 'F' is the feet of the filament. The latitude of 'F' and '6' is same and the longitude separation is 8.7$^\circ$, this is used to determine filament inclination along red arrow in section 2.3.4.}
\label{fig:height}
\end{figure}

 The height of the filament spine during the eruption at 10:56 UT was reconstructed by using a routine called SCC\_MEASURE, which is a part of SolarSoft IDL library for SECCHI data analysis \cite{Thompson06}. The
routine measures 3D coordinates from two STEREO images. It is a widget based application that allows a user to select with the cursor a
common feature in both images.  The user identifies a point on one image and then selects the point corresponding to the same feature on another image.  The 3D coordinates are then calculated as longitude, latitude and radial distance from the center of the Sun. The radial distance is used to measure height above the solar surface. Figure~\ref{fig:height} gives the various locations on filament spine for which the height is measured by above-mentioned procedure. The longitude, latitude and height of these locations is given in Table 2. Note that the location marked by  `F' is the feet of the filament.

 It must be mentioned that this technique relies on identification of identical features in STEREO A and B, which is difficult due to very different projections in our case (separation angle 52.4 degrees). However, in SCC\_MEASURE routine  we first select a point on the filament spine in STEREO-A, and immediately the epipolar line is displayed in the adjacent STEREO-B image. Then we have to select a point along this epipolar line where it intersects the filament spine as viewed in STEREO-B image.  This removes possibility of  error in locating same feature in one direction (vertical direction), this is called ''epipolar constraint" \cite{Inhester06}. However, the main source of error that remains is the error in identification of the filament spine in STEREO-B, as it appears quite differently in both images. An estimate of error in determination of height using this technique was given by \inlinecite{Liewer09}. They relate the error in locating the common feature in the image to an error in height by the relation $\Delta h~ \approx ~ \Delta x / sin \phi$, where $\Delta x$ is the error in locating common feature in pixels and $\phi$ is the separation angle of the two spacecraft. During our observations separation angle was about 52.4$^\circ$, which leads to an error in height determination to about $\Delta h /R_{sun} \approx 0.12\%$, taking  $R_{sun}$ for SECCHI/EUVI pair as $\approx$ 1001 pixels.

In Figure~\ref{fig:height} we can see that except location 2, other locations marked 1, 3, 4, 5 and 6 are clearly identified, thanks to the sharp contrast of the filament spine. The contrast is good enough so as to  determine identical feature in both images within 2 to 3 pixels. Taking $\pm$ 3 pixels as the worst case error, this translates to an error of $\approx \pm 2.5 Mm$ in the determination of height of locations 1, 3, 4, 5 and 6. For location 2, we take worst error to be $\pm$ 6 pixels (filament width at location 2 in STEREO-B), which translates to an height error of $\approx \pm 5 Mm$. These errors are small as compared to the heights determined in Table 2.

However, one must keep in mind the inherent limitation of line-of-sight integration effect while interpreting  reconstructed filaments using EUVI data. Features may look very different from different viewing angles and background features always add to the confusion. The use of scc\_measure has been possible only during the eruption when identical structures  could be identified in STEREO A and B images. Nevertheless, these reconstructions are the best that can be done and give more information than hitherto studied non-stereoscopic images.

\subsubsection{Inclination estimate using SCC\_MEASURE}
\begin{figure}    
\centerline{\includegraphics[width=.25\textwidth,clip=]{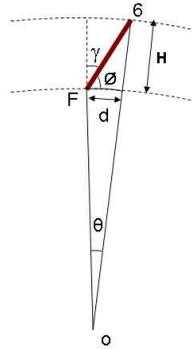}}
\caption{  The filament inclination is estimated using the 3-D coordinates of locations 'F' and '6' derived in Table 2. The latitude is same, while longitude separation $\theta = 8.7^\circ $, height $H=99 Mm$, solar radius $R_{sun}=700Mm$, $d=\theta R_{sun} = 106 Mm$. Thus, using $tan \phi = H/d$, we get $\phi=43^\circ$ and $\gamma =47^\circ$.}
\label{fig:inclin}
\end{figure}

 The three dimensional spherical coordinates of the filament determined in Table 2 above are used to calculate the filament inclination $\gamma$.
The two positions marked 'F', the filament feet, and '6' on the filament spine are used to determine the inclination of filament sheet. These two points are chosen because they correspond to same latitude. Thus, choosing these  points will yield filament inclination along latitudinal direction. Further, these locations have longitudinal separation, $\theta = 8.7^\circ$. Figure~\ref{fig:inclin} illustrates that  using the height, $H$, of '6' to be 99 Mm, and $\theta =8.7^\circ$ we can estimate $d=\theta R_{sun}=$106 Mm and thus determine angles $\phi =atan(H/d) $ to be 43$^\circ$, and thus the inclination, $\gamma$, of the segment of the filament sheet shown by red arrow between 'F' and '6', is about 47$^\circ$.

\begin{table}
\caption{3-D coordinates of the filament spine determined using SCC\_MEASURE routine}
\label{tab:2}
\begin{tabular}{lccr}
\hline
Location & Longitude & Latitude & Height (Mm)  \\
\hline
1& -12.6 &-22.4 &50 \\
2&  -9.4&-27.6 &63 \\
3&  2.4 & -24.2&116 \\
4&  5.3& -21.9&123 \\
5&  7.5& -19.6&113 \\
6&  9.0 & -18.3&99 \\
F&  0.3 & -18.3 & 0 \\
\hline
\end{tabular}
\end{table}

\section{Discussion and Conclusions}

A long  S-shaped filament composed of several segments and located in the Southern hemisphere was disappearing between May 20-22, 2008. This was observed during a coordinated campaign (JOP-178) involving ground-based instruments, THEMIS on the Canary Islands, MSDP at the Meudon solar tower, GONG magnetograph and H$\alpha$ telescope in Udaipur, as well as space-based instruments SOHO/MDI and SECCHI/EUVI aboard STEREO. H$\alpha$ instruments observed the progressive disappearance of the filament, segment after segment, between May 20 to May 22.  It was a long process with some impulsive phases. The last H$\alpha$ segment was visible  on May 22 till 10:00 UT while the long spine of the S-shaped filament was observable in He II 304 \AA\ till 15:30 UT. Before disappearing filament segments showed high dynamics in H$\alpha$ with blue and red shifts parallel to the filament axis.

MDI and GONG magnetograms show that the filament is located along the polarity inversion line between two weak magnetic field regions. The magnetic field in the filament channel was weak and rapidly changing. Local correlation tracking techniques applied to GONG polarities showed strong shear between these two regions two days before the eruption, then some vortex pattern and no noticeable motion after the eruption.
THEMIS magnetograms allowed us to identify weak  minority polarities associated with the filament feet. The canceling flux/polarities in the vicinity of the feet were observed in the GONG movies a few hours before the disappearance of the associated segment.  Such cancelation of flux and disappearance of the foot-point barbs where the filament is tied to the photosphere could lead to  eruptions (Raadu {\it et al.} 1987,1988).

 The angular separation  between STEREO A and STEREO B spacecraft was 52.4 degrees during our observations, which gave quite different views of the filament. We used different methods to study the filament disappearance using He II 304 filtergrams: (i) study of filament dynamics and estimation of the projected rise velocity of thread-like structures using He II 304 movies, (ii)  by overlaying a spherical grid over the solar surface drawn over a sphere of 700 Mm to identify elevated features and filament feet and put an upper limit on the filament inclination, and (iii) using the triangulation algorithm SCC\_MEASURE (part of STEREO data analysis library in Solarsoft package) to compute the altitude and inclination of the filament over the chromosphere during its disappearance phase. These  methods give consistent results.

 The stereoscopic reconstructions using SECCHI/EUVI observations of 19 May 2007 filament eruption event were reported recently by \cite{Liewer09} and \cite{2008SoPh..252..397G}. While \inlinecite{Liewer09} used scc\_measure for reconstruction, \inlinecite{2008SoPh..252..397G} used optical flow method to find displacements between features in stereoscopic pairs of SECCHI/EUVI 304 \AA\ images. The results of these two techniques are in good agreement \cite{Liewer09}. These  results for the 19 May 2007 event showed that the filament eruption was  asymmetric and whip-like. The filament disappearance in our case is different from their study in two ways (i) filament was rooted in a weak diffuse bipolar magnetic region not in an active region, and (ii) there was no CME recorded for our filament disappearance event (however, one cannot rule out a CME as it could be below detection threshold). The length of the filament is too large spanning more than 10$^\circ$ in latitude, as seen in He II 304 \AA\ images. We also use scc\_measure to derive three-dimensional geometry of the filament which is used to derive its height and inclination. The filament was highly inclined to solar normal by about 47$^\circ$. Such large inclinations are interesting as  theoretical models of filaments consider filament sheet as thin vertical slab.  The maximum filament height determined during the disappearance phase around 10:56 UT is estimated to be about 123 Mm in the middle portion of the filament. Determining height evolution of filament during the disappearance phase was difficult as it became quite diffuse, making identification of common features very difficult. The presence of fine threads running parallel to filament spine could be seen in movies. These threads are so tiny that it was difficult to track them over the chromosphere in the later phases, i.e., after around 11:00 UT. Only the EUVI movies allowed us to distinguish them over the chromosphere. The initial dense filament became an  untwisting flux rope with multiple threads with a fan-shaped structure which rose and disappeared one by one into the corona. The plasma becomes optically so thin as the flux rope rapidly expands in the corona that it remains no longer visible.

 Bright structure observed in STEREO A and B images at 195 \AA\ is interpreted as reconnection loop system, located below the filament (flux-rope).  This bright structure is not visible earlier at 00:00 UT and appears during the onset of rapid disappearance phase at about 06:00 UT. The plasma of the filament is less and less dense as the flux rope  rises and expands. As the plasma is dispersed it is no more visible in filter 304 \AA\  nor in 195 \AA\.~ No CME was reported during this last phase (LASCO was not observing) by COR1 and COR2, the two coronagraphs aboard STEREO. The CME might be too faint to be detected or the magnetic field might be too weak to prevent the plasma's expansion to undetectable density levels.

 Ground based data are very useful to understand the context of the event. Up to now we have the knowledge of eruption phenomena principally by using instruments with low temporal and/or spatial resolution. These observations are still largely used to study the association of filament eruption and CMEs. On the other hand high-cadence instrumentation with high-resolution have made a breakthrough in the nature of the dynamics of filament fine-structures. Non-eruptive filaments show counter-streaming flows along the fine structures and up-and-down flows in the filament-end (\cite{Zirker98}, \cite{Lin2005}, \cite{Berger2008}). These velocities are lower than 10 km s$^{-1}$. With STEREO and ground-based observations we can find some relationship between the dynamics at large-scale and at small-scale. The H$\alpha$ doppler velocities associated with the filament, that we measure in the present paper, are relatively small but they confirm previous observations of activated filaments with low spatial-resolution instruments (\cite{Schmieder85a},\cite{Schmieder85b}). The pair of aligned and elongated regions of oppositely directed velocities are interpreted in terms of a twisted  magnetic flux rope. The STEREO observations allow perspective effect to be removed in doppler velocities by taking into account the filament inclination. The corrected velocities are about 1.5 times more. The Doppler-shifts are also lower due to the computation by using the bisector method and not taking account the background chromosphere. In the paper by \cite{Schmieder85a} and \cite{Schmieder85b}, they could derive larger velocities by a factor 2 between the foot-points and of the order of 10 km s$^{-1}$ in the feet using a cloud model method. That is probably what we could expect with our present observations using the cloud model method. The standard method (bisector) indicates the general trend of the velocities with large uncertainty but indicates clearly that the filament is activated. The day before we did not observe such organized velocity pattern.

In the future, development of ground based instruments like dual-beam doppler imaging system, being developed at Udaipur Solar Observatory \cite{Joshi2009} for detecting filament activation using high-cadence H$\alpha$ dopplergrams,  and combined observations with STEREO will lead to further developments in our understanding of the filament eruptions.

\begin{acks}
 The authors thank CNRS for allocating observing time on THEMIS. THEMIS is a French-Italian telescope installed at Observatorio El Teide, Tenerife, Spain. The observations from BBSO, GONG and SOHO are acknowledged. We thank V. Bommier for providing her UNNOFIT code for inversion of Stokes profiles. We also thank Arturo Lopez Ariste, G. Ruyman and  Cyril Delaigue for help in observations and data reductions. Further, SG acknowledges CEFIPRA funding for his visit to  Observatoire de Paris, Meudon, France under its project No. 3704-1. Also, SG acknowledges travel support to THEMIS, Tenerife by Observatoire de Paris, Meudon. We thank Fr$\acute{\rm e}$d$\acute{\rm e}$rique Auch$\grave{\rm e}$re for the STEREO movies. We also thank G. Molodij, J. Moity for the observations at the solar tower and P. Mein for helping in the MSDP data reduction. This work was supported by the European network SOLAIRE (MTRN\_CT\_2006\_035484).
\end{acks}

\bibliographystyle{spr-mp-sola}
%

\end{article}

\end{document}